\documentclass[twocolumn,aps,superscriptaddress]{revtex4-1}

\usepackage{dcolumn}
\usepackage{graphicx}
\usepackage{graphics}
\usepackage{dcolumn}
\usepackage{bm}
\usepackage{bbm}
\usepackage{amsfonts}
\usepackage{amsmath}
\usepackage{amssymb}
\usepackage{dsfont}
\usepackage{mathrsfs}
\DeclareMathAlphabet\mathbfcal{OMS}{cmsy}{b}{n}

\newcommand\redout{\bgroup\markoverwith{\textcolor{red}{\rule[.5ex]{2pt}{0.4pt}}}\ULon}

\usepackage[colorlinks=true,citecolor=blue]{hyperref}
\hypersetup{colorlinks=true,citecolor=blue,linkcolor=red,urlcolor=blue}

\begin{document}

\title{Anisotropic RKKY interactions mediated by $j=3/2$ quasiparticles\\ in half-Heusler topological semimetal}

\author{Reza G. Mohammadi}
\address{Department of Physics, Institute for Advanced Studies in Basic Sciences (IASBS), Zanjan 45137-66731, Iran}

\author{Ali G. Moghaddam}\email{agorbanz@iasbs.ac.ir}
\address{Department of Physics, Institute for Advanced Studies in Basic Sciences (IASBS), Zanjan 45137-66731, Iran}
\address{Research Center for Basic Sciences \& Modern Technologies (RBST), Institute for Advanced Studies in Basic Science (IASBS), Zanjan 45137-66731, Iran}

\begin{abstract} 
We theoretically explore the RKKY interaction mediated by spin-3/2 quasiparticles in half-Heusler topological semimetals in quasi-two-dimensional geometries. 
We find that while the Kohn-Luttinger terms gives rise to generalized Heisenberg coupling of the form ${\cal H}_{\rm RKKY} \propto {\sigma}_{1,i} {\cal I}_{ij} {\sigma}_{2,j}$ with a symmetric matrix ${\cal I}_{ij}$, addition of small antisymmetric linear spin-orbit coupling term leads to Dzyaloshinskii-Moriya (DM) coupling with an antisymmetric matrix ${\cal I}'_{ij}$. We demonstrate that besides the oscillatory dependence on the distance, all coupling strengths strongly depend on the relative orientation of the two impurities with respect to the lattice. This yields a strongly anisotropic behavior for ${\cal I}_{ij}$ such that by only rotating one impurity around another at a constant distance, we can see further oscillations of the RKKY couplings. This unprecedented effect is unique to our system which combines spin-orbit coupling with strongly anisotropic Fermi surfaces. We further find that all of the RKKY terms have two common features: a tetragonal warping in their map of spatial variations, and a complex beating pattern. Intriguingly, all these features survive in all dopings and we see them in both electron- and hole-doped cases. In addition, due to the lower dimensionality combined with the effects of different spin-orbit couplings, we see that only one symmetric off-diagonal term, ${\cal I}_{xy}$ and two DM components ${\cal I}'_{xz}$ and ${\cal I}'_{yz}$ are nonvanishing, while the remaining three off-diagonal components are identically zero. This manifests another drastic difference of RKKY interaction in half-Heusler topological semimetals compared to the electronic systems with spin-1/2 effective description.
\end{abstract}

\maketitle
\section{Introduction}
Over the past decades, predictions and subsequent realizations of topological phases of matter have revolutionized our understanding of condensed-matter systems and electronic structure of materials \cite{haldane-nobel}. 
Time-reversal invariant (TRI) topological insulators (TIs) in two and three dimensions (2D and 3D), topological superconductors exhibiting exotic Majorana zero modes (MZMs) at the boundaries, and Weyl/Dirac semimetals are among well-known examples which possess a nontrivial topology in their low-energy band structures \cite{kane-rmp,zhang-rmp,vishwanath-rmp2018}. 
In many topological phases, spin-orbit coupling (SOC) is a key player which can give rise to topological phase transition by inverting the order of low-lying electronic bands. 
The most famous examples of the TRI topological insulators derived by SOC are HgTe/CdTe quantum wells in 2D and Bi/Sb binary alloys and compounds in 3D \cite{koenig-science2007,hsieh-nature2008,hsieh2009tunable,zhang2009TRI,chen-science2009,xia2009observation}.
Very intriguingly, the band inversion driven by strong SOC 
has been predicted to take place in various ternary half-Heusler compounds containing rare-earth elements like Lanthanum and Yttrium. When the high cubic symmetry is preserved, these materials are topological semimetals (TSM) rather than insulators or semiconductors, but they can be turned to TIs by distorting the cubic symmetry  \cite{felser2010natmat,hasan2010natmat,xiao2010halfheusler}.
The low-lying electronic excitations in these semimetals consist of $\Gamma_8$ bands having a total angular momentum of $j=3/2$ \cite{dresselhaus}. Furthermore, most of the half-Heusler systems show correlated symmetry-broken ground states including superconducting and antiferromagnetic phases \cite{yptbi-sup-topo-prb2011,topo-sup-luptbi-prb2013}. Of particular interest, higher spin of low-energy quasiparticles lead to further Cooper pairing channels particularly the so-called $j=3$ septet pairing \cite{brydon-prl,lee-prb-2017,kimeaao-sci-adv,timm-17,fu-lee-prx,ghorashi-prb19,moghaddam14}. 
\par
In spite of studies about the magnetic properties of spin-orbit coupled half-Heusler systems, which suggest some applications in spintronics as well, the role of their low-energy spectrum has been overlooked so far \cite{felser2018heusler,felser-rev-energy}. 
So a natural question that arises here is how the magnetic
features in the half-Heusler TSMs can be influenced by the low-energy excitations 
possessing an effective total angular momentum of $j=3/2$
and various SOC terms. 
We must remind in spite of the fact that the magnetic orderings mainly originates from the localized electronic states, there is a range of rich physical phenomena where low-lying electronic states can influence the magnetic properties \cite{kubler2017book}. One of the key phenomena in this context is the well-known Ruderman-Kittel-
Kasuya-Yosida (RKKY) interaction or indirect exchange coupling between localized magnetic moments mediated by conducting electrons \cite{kittel,kasuya,yosida}.
It is known that this interaction plays an essential role in inducing various types of magnetic phases in metallic and semiconducting electronic systems at the presence of magnetic adatoms \cite{ohno-rmp}. Particularly, RKKY coupling of localized magnetic moments has been extensively assumed as the key mechanism to induce ferromagnetism in dilute magnetic semiconductors \cite{ohno-science,ohno-prb}. Very recently similar predictions and also experiments have been made for magnetically doped TIs in which more exotic phases like a disordered spin-glass emerges besides ordered ferromagnetic state \cite{abanin,rkky-topological-zhang-2009,chang2011,hor-magnetic-TI,checkelsky2012dirac}. Another interesting feature of RKKY interaction is its strong dependence on the dimension of the host electronic system \cite{rkky-1d-yafet,rkky-2d-1987,rkky-1d-2d-dugaev,rkky-prb-any-d}, and the dispersion relation of the low-energy excitations \cite{rkky-graphene-2010,rkky-graphene-2011,rkky-kogan,rkky-weyl,hosseini,rkky-akbari}.
In the case of a clean three dimensional electron gas, RKKY interaction has an isotropic Heisenberg-like form with a coupling strength which shows oscillatory decaying dependence on the distance of two localized moments. However, in confined geometries and lower dimensions as well as in the presence of SOC, the properties of RKKY coupling can drastically change. It has been found that Rashba SOC in electron gas leads to anisotropic indirect exchange coupling between the magnetic impurities. As a result, extra forms of RKKY interactions including the Dzyaloshinskii-Moriya (DM) and Ising terms exist \cite{bruno-prb-2004,chao2007,loss2008,egger2009,jiaji2010,loss2013,rkky-shiba-3d,tse-bulk-rashba,saha2019}.   
\par
In this work, motivated by the aforementioned question, we investigate the RKKY interaction between magnetic impurities in TSMs with effective spin-$3/2$ quasiparticles. This problem has not been explored so far and can pave a way to induce and tailor magnetic ordering in TSMs.
%described by the four-band Luttinger-Kohn Hamiltonian in addition of an antisymmetric SOC (ASOC) term. 
%This problem is important in finding a route to tailor magnetic ordering in TSMs and may lead to spintronic applications. 
Furthermore, since there are various suggestions for inducing topological superconductivity and MZMs in RKKY systems coupled to superconductors \cite{loss-mzm,simon-2013,glazman-zmz,sarma-mzm,ojanen}, our study can trigger exploration of MZMs in half-Heusler TSMs specially those with intrinsic superconductivity. Here, concentrating on quasi-two-dimensional structures, we find that both the spatial dependence of the RKKY coupling constants and their spin structure have strong anisotropies. The anisotropy in the spatial dependence of RKKY couplings mainly originates from the strongly anisotropic low-energy dispersion of electronic excitations. However, the anisotropy in the matrix structure of RKKY coupling can be ascribed to the low dimensionality of the system and more importantly the presence of different SOC terms which are closely related to the spin-$3/2$ nature of the quasiparticles. Also, the direction dependence of the Fermi wavelength gives rise to a beating pattern in spatial variations which are essentially different from those in electronic systems with two or more Fermi surfaces. On the other hand, the presence of antisymmetric SOC (ASOC) with linear momentum dependence leads to the existence of DM-type RKKY interaction. Since the anisotropy of the dispersion relation and the tetragonal warping effects do not disappear even at very low dopings and close to the band touching point, all main features of the RKKY coupling are qualitatively maintained irrespective of the Fermi energy of the system.  
Concentrating on YPtBi as the prototype $j= 3/2$ TSM, we find that in spite of qualitative similarities, some quantitative differences between hole- and electron-doped cases can be seen. These findings, besides providing significant differences with previously studies spin-orbit coupled systems, can have intriguing consequences for the ordered phases of magnetically-doped TSMs and also possible realization of MZMs in these systems.
\par
In the remainder of the paper, we first introduce the low-energy Hamiltonian of the topological semimetals and the formalism of calculating indirect exchange coupling for systems with strongly anisotropic band structures (Sec. \ref{sec-2}). Then in Sec. \ref{sec-res} we present the results complemented by the discussion over them. The paper is ended up with the concluding remarks in Sec. \ref{conc}

%%%%%%%%%%%%%%%%%%%%%%%%%%%%%%%%%%%%%%%%%%%%%%%%%%%
%%%%%%%%%                              MODEL SECTION                     %%%%%%%%%%%%%%%%%
%%%%%%%%%%%%%%%%%%%%%%%%%%%%%%%%%%%%%%%%%%%%%%%%%%%

\section{Model and basic formalism}
\label{sec-2}
\subsection{Effective Hamiltonian of TSM}
\label{subsec-a}
We start with the low-energy effective Hamiltonian
\begin{eqnarray}
\label{HTSM}
\check{\cal H}_{\rm TSM}
=\check{\cal H}^{(0)} + \check{\cal H}^{(1)}, 
\end{eqnarray}
at the vicinity of the ${\rm \Gamma}$ point for the half-Heusler semimetals, which consists of two parts
\begin{eqnarray}\label{luttinger-kohn}
&&\check{\cal H}^{(0)}=\alpha_1 |{\bf k}|^2\check{\mathbbm{1}}+ \alpha_2 ({\bf k}\cdot \check{\mathbfcal J})^2+\alpha_3 \sum_{i=1}^{3}k_i^2 \check{\cal J}_i^2,\\ \label{asoc}
&&\check{\cal H}^{(1)}=\alpha_4 {\bf k}\cdot \check{\mathbfcal T}~,~~~\check{\cal T}_i=\{\check{\cal J}_i,\check{\cal J}^2_{i+1}-\check{\cal J}^2_{i+2} \},
\end{eqnarray}
corresponding to the Luttinger-Kohn (LK) Hamiltonian and ASOC terms, respectively.
Here, $\check{\mathbfcal J}=\check{\cal J}_i \hat{\bf e}_i$ denotes the $j=3/2$ angular momentum operators having $4\times 4$ matrix representation, ${\bf k}$ is the momentum of the excitations, and $\{ \check{\cal A},\check{\cal B} \}$ indicates the anti-commutator of two operators $\check{\cal A}$ and $\check{\cal B}$. The coefficients $\alpha_i$ are material dependent and can be extracted by fitting the low-energy model with the results of \emph{ab initio} calculations for a certain topological semimetallic half-Heusler material like YPtBi and LuPtBi \cite{brydon-prl}. The first term of the LK Hamiltonian corresponds to the spin-independent mass term and the other two terms represent the symmetric SOC (SSOC) which both of them are quadratic in momentum and the spin-$3/2$ operators $\check{\cal J}_i $. The difference between second and third terms can be elucidated from the point group symmetry perspective which states while the second term have full spherical symmetry, the third term has a reduced symmetry corresponding to a cubic lattice. Intriguingly, the existence of the SSOC terms is deeply connected to the higher spin of LK Hamiltonian and the algebraic structure of spin-$3/2$ operators. While in the case of systems with effective spin-$1/2$ and represented by Pauli matrices $\sigma_i$, only ASOC terms appear which are linear in spin operators and contain only odd powers of momentum. In this respect, the LK Hamiltonian and spin-$3/2$ TSM are fundamentally different from the whole variety of effective spin-$1/2$ systems including those with linear and cubic Rahba SOCs as well as Weyl/Dirac semimetals. In half-Heusler TSMs, both SSOC terms are very strong and their corresponding coefficients $\alpha_2$ and $\alpha_3$ are in the same order as $\alpha_1$. On the other hand, the ASOC term (\ref{asoc}) which is present due to the inversion
symmetry broken tetrahedral crystal field, acts as a small correction to the LK Hamiltonian \cite{lee-prb-2017}. In spite of its very different matrix structure containing third powers of spin-$3/2$ operators, ASOC term (\ref{asoc}) shares some similarities with the simple Rashba term $\cal{H}_{R}\propto \hat{\bf z}.({\bf k}\times {\bm \sigma})$ particularly because of their linear momentum-dependence. In the next parts, we will see that how these features yield interesting results for the RKKY physics in half-Heusler TSMs particularly compared to other electronic systems. 
\par
In order to proceed, we make a couple of assumptions which simplify mathematical treatment without losing the generality of the problem or any significant physics at least in a qualitative manner. First, we treat $\check{\cal H}^{(1)}$ as a perturbation, provided by the fact that it has a weaker effect in comparison with other terms when we stay far enough from the neutrality or band touching point. By considering the first term in LK Hamiltonian as the dominant term, we can estimate that only for energies below a certain value ($|\varepsilon| \lesssim \alpha_4^2/\alpha_1$), the ASOC term has comparable or stronger effect than LK Hamiltonian. Hence, for the prototype half-Heusler material
YPtBi with $\alpha_{1}\sim 20\,(a_{0}/\pi)^2$eV and $\alpha_4 \sim0.1\,(a_{0}/\pi)$eV, for energies quite higher than $\alpha_4^2/\alpha_1\sim 0.5\,$meV, the effect of ASOC term in the dispersion becomes small such that it can be treated perturbatively.
\par
As a second assumption in our study, we concentrate on effectively 2D structures which can be realized by considering thin films of TSMs. Therefore, we can directly compare our results for RKKY interaction with those in 2D electron gas at the presence of SOC or other related systems like topological insulators, besides the possibility of implicit comparison with 3D Rashba systems, and Dirac/Weyl semimetals. In order to deduce an effective 2D Hamiltonian, we use the approximation of setting momentum in the perpendicular direction to zero ($k_z\to 0$) except for 
the nonvanishing quantum mechanical expectation value of the quadratic term $\langle k_z^2\rangle \sim \pi^2/d^2$ where $d$ is the width of thin film. So we arrive in the 2D analogous of Hamiltonians (\ref{luttinger-kohn}) and (\ref{asoc}) having only $k_x$ and $k_y$ terms, besides an additional term $\alpha_1 \langle k_z^2\rangle \check{\mathbbm{1}} + (\alpha_2+\alpha_3)  \langle k_z^2\rangle \check{\cal J}_z^2 $ attributed to the quantum confinement. Since the first part of the extra terms which is indeed a constant energy shift can be simply absorbed in the chemical potential $\mu$, we only need to consider an explicit term $\beta \check{\cal J}_z^2$ with $\beta\equiv (\alpha_2+\alpha_3)  \langle k_z^2\rangle$. 

\subsection{RKKY interaction and Green's functions}
\label{subsec-b}
For studying the indirect exchange interaction between magnetic impurities, we consider them as localized classical moments (denoted by ${\bm \sigma}=\sigma_i \hat{\bf e}_i$), similar to the original RKKY problem. Then we assume that the localized spins are coupled to the low-lying spin-$3/2$ electronic excitations inside TSM
by a $s-d$ type Hamiltonian 
\begin{equation}\label{s-d}
{\cal H}_{\rm imp}=-\lambda \sum_\alpha\check{\mathbfcal J} \cdot {\bm \sigma}_\alpha  \delta  ({\bf r}-{\bf R}_\alpha),
\end{equation}
in which $\lambda$ is the exchange coupling energy between the localized moments and the spin of delocalized charge carriers.
In principle because of the large spin of quasiparticles ($j=3/2$), one can think about more complex model Hamiltonians for coupling of a magnetic impurity with electronic excitations of the host material. 
For instance, assuming heavy metal impurities from $4f$ elements, the exchange interaction between the impurity and the electronic excitations are described by the $s-f$ exchange interaction. In these models, depending on the strength of spin-orbit coupling (${\mathbf S}\cdot {\mathbf L}$) of the impurity, the single electron spin operator ${\bm \sigma}_{\alpha}$ is replaced by the total spin ${\bf S}_{\rm imp}$ or total angular momentum ${\bf J}_{\rm imp}$ of the impurity, but the interaction constant $\lambda$ is assumed to remain a single constant. On the other hand, 
starting from the Anderson Hamiltonian, interactions beyond conventional $s-d$ and $s-f$ exchange couplings are also possible, in which the change in the magnetic quantum numbers of the localized moment is not limited to $0,\pm1$ as first shown by Coqblin and Schrieffer \cite{schrieffer69}. They have also found that RKKY coupling between localized moments with large ${\bf J}_{\rm imp}$ radically changes in their model,
while in the framework of conventional $s-f$ model the form of RKKY interactions remains unchanged. Therefore, provided by the fact that the magnetic impurities in RKKY problem are essentially classical spins rather than quantum mechanical quantities, we stick to the simple extension of conventional $s-d$ coupling as introduced above. Remarkably and in accordance with Ref. \cite{schrieffer69}, this treatment unifies $s-d$ and $s-f$ exchange models and we only need to replace the impurity spin operators ${\bm \sigma}$ with ${\bf S}_{\rm imp}$ or ${\bf J}_{\rm imp}$ in the RKKY interaction formula presented below. Nevertheless, generalization to the cases like Coqblin-Schrieffer model which are more relevant in studying strong correlations physics of the impurities like Kondo problem, falls beyond the scope of the present work and is left for future studies.

%Therefore, because of the fact that in RKKY interaction the magnetic impurities are treated as classical spins rather than quantum mechanical quantities, we stick to the simple extension of conventional $s-d$ coupling as introduced above. 
Now, applying the standard second order perturbation theory with respect to the impurity Hamiltonian (\ref{s-d}), we arrive in the RKKY interaction between two localized spins positioned in ${\bf R}_1$ and ${\bf R}_2$, as 
\begin{equation}\label{rkky}
{\cal H}_{\rm RKKY}=\lambda^2 {\bm \sigma}_1 \cdot {\bm \chi}({\bf R}_1-{\bf R}_2) \cdot {\bm \sigma}_2,
\end{equation}
in which ${\bm \chi}({\bf R})$ is a $3\times 3$ matrix representing the spin susceptibility of the spin-$3/2$ TSM.
The components of ${\bm \chi}$ are given by
\begin{eqnarray}
&&{ \chi}_{ij}({\bf R}) = \Im
\int_{-\infty}^{\varepsilon_F}  \frac{d\omega}{\pi}  \,{\rm Tr}\left[\check{\cal J}_i \check{G}_\omega({\bf R})  \check{\cal J}_j \check{G}_\omega(-{\bf R})\right],
\\
&& \label{green-fourier}
\check{G}_\omega({\bf R})=\int    \frac{ d^d{\bf k} }{(2\pi)^{d} }~ e^{i{\bf k}\cdot {\bf R}} \check{G}_\omega({\bf k}) . 
\end{eqnarray}
in which $\check{G}_\omega({\bf k})$ denotes the momentum-space representation of the Green's function corresponding to the Hamiltonian (\ref{HTSM}) governing the low-energy excitations of the TSM. Here, $\varepsilon_F$ is the Fermi energy measured with respect to the band touching point and $\Im$ gives the imaginary part.
\par
\begin{figure}
    \centering\includegraphics[width=.65 \linewidth]{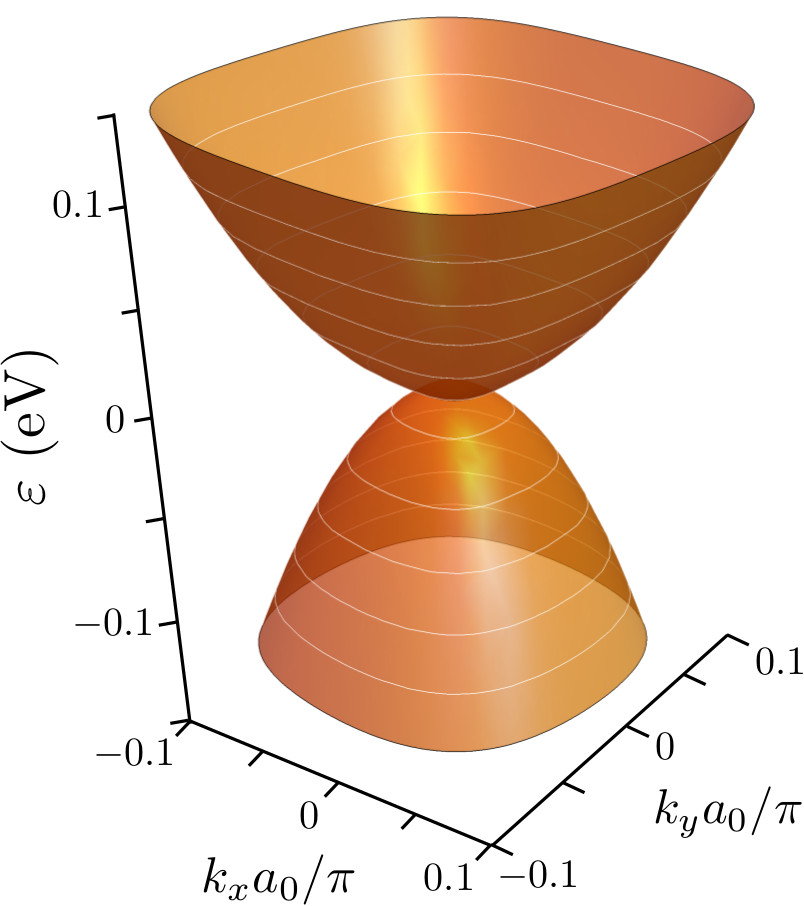}
    \caption{(Color online)
Low-energy dispersion of a half-Heusler TSM near the ${\rm \Gamma}$ point, described by LK effective Hamiltonian. Considering YPtBi as an example, the Hamiltonian parameters used for this plot are $\alpha_{1}=20 (a_{0}/\pi)^2$eV, $\alpha_{2}=-15 (a_{0}/\pi)^2$eV, $\alpha_{3}=-5 (a_{0}/\pi)^2$eV. Difference in the effective masses of conduction and valence (electron and hole) bands and tetragonal warping of isoenergy lines in all energies are two important features of the band structure which can be easily seen in this plot.}\label{fig0}
\end{figure}

Treating the ASOC term as a small perturbation, the Green's function for the 2D version of Hamiltonian (\ref{HTSM}) reads
\begin{equation}
\check{G}_\omega({\bf k}) \approx \check{G}^{(0)}_\omega({\bf k}) + 
\check{G}^{(0)}_\omega({\bf k}) \check{\cal H}^{(1)} \check{G}^{(0)}_\omega({\bf k}).
\label{green-0th-1st}
\end{equation}
The zeroth order Green's function $\check{G}^{(0)}_\omega({\bf k})=(\omega - \check{\cal H}^{(0)})^{-1}$ of the 2D LK Hamiltonian (without ASOC term) is given by
\begin{eqnarray}
&&\check{G}^{(0)}_\omega({\bf k})=\dfrac{1}{\mathcal{D}}
\left\{
\left[\omega- \left( \alpha_{1}+\frac{5}{2} \alpha_2 +\frac{5}{2} \alpha_3\right) k^{2} \right]\check{\mathbbm{1}}
\right.
\nonumber\\
  &&~~~~+ \left . 
\alpha_2 ({\bf k}\cdot \check{\mathbfcal J})^2+\alpha_3 \sum_{i=1}^{2}k_i^2 \check{\cal J}_i^2
+ \left(  \check{\cal J}_{z}^2-\frac{5}{2}   \right)\beta
\right \} , \label{green-2d} 
\end{eqnarray}
with
\begin{eqnarray}
{\mathcal{D}}&=&k^4\left[
\alpha_1^2+\frac{5}{2} \alpha_1 ( \alpha_2 +\alpha_3 )+ \frac{9}{16} ( \alpha_2 +  \alpha_3 )^2\right]
\nonumber
\\
&+&\frac{k^2}{8} \left[
( \alpha_2+\alpha_3 ) (33 \beta -20 \omega )+4 \alpha_1 (5 \beta -4 \omega )
\right]
\nonumber \\
&+&\frac{9 \beta ^2}{16}-\frac{5 \beta  \omega }{2}+\omega ^2+3  \alpha_3  (2 \alpha_2 + \alpha_3 ) k_x^2 k_y^2.
\label{d}
\end{eqnarray}
Accordingly, the eigenvalues of the 2D correspondence of Hamiltonian (\ref{luttinger-kohn}) when $\beta=0$ are also obtained as
\begin{eqnarray}
&&\varepsilon^{(0)}_{\pm}(k,\theta_k)=k^2\big[
\alpha_1+\frac{5}{4}(\alpha_2+\alpha_3)\big]
\nonumber\\
&&~~~~~~~~~~ \pm k^2 
\sqrt{\alpha_2^2+(\alpha_3^2+2\alpha_2\alpha_3)\frac{5+3\cos (4\theta_k)}{8}}\,
,~~~\label{eigens}
\end{eqnarray}
with $\theta_k=\arctan(k_y/k_x)$. Both energy bands $\varepsilon^{(0)}_{\pm}(k,\theta_k)$ are two-fold degenerate, but the degeneracy is slightly lifted when the ASOC is taken into account. The energy dispersion 
(\ref{eigens}) clearly shows that the anisotropy of the band structure which originates from $\alpha_3$-term, persists in all energies and gives rise to a tetragonal warping of Fermi surfaces. Furthermore, one can see that the two branches $\varepsilon^{(0)}_{\pm}$ have opposite signs for their effective masses (corresponding to electron and hole bands) when $\alpha_1$ lies in the range between $-(1/4)(\alpha_2+\alpha_3)$ and $-(9/4)(\alpha_2+\alpha_3)$. These are the situations where the effective Hamiltonian corresponds to TSM. Otherwise, when $\alpha_1$ is out of this range, both branches will be electron- or hole-like where in the second case, $\varepsilon^{(0)}_{\pm}$ correspond to heavy-hole and light-hole bands of the typical semiconductors. To better illustrate the aforementioned aspects of the TSM band structure, Fig. \ref{fig0} shows the low-energy dispersion of YPtBi confined in a quasi-2D geometry and given by KL Hamiltonian. 
\subsection{Angular harmonics expansion}
In this subsection, we present a framework of expansion over the angular harmonics $\psi_m(\phi)=e^{im\phi}$, for the spatial dependence of the spin susceptibility and RKKY coupling constants. The framework is specially useful for the systems with anisotropic dispersion relations like $\check{\cal H}_{\rm TSM}$. In 3D cases a similar formulation can be developed by replacing $\psi_m(\phi)$ with the spherical harmonics $Y_{lm}(\theta,\phi)$.
\par
We start by writing the Fourier relation (\ref{green-fourier}) for the Green's function in two dimensions. By choosing polar coordinates ${\bf k}=k(\cos\theta_k,\sin\theta_k)$ this leads to
\begin{eqnarray}
&&\check{G}_\omega({\bf R}) =
\int \frac{d^2{\bf k}}{(2\pi)^{2}}~ e^{i{k}{R}\cos{(\theta_{k}-\phi)}} \check{G}_\omega({\bf k}),
\end{eqnarray}
in which $d^2{\bf k}\equiv kdkd\theta_{k}$. The polar representation of the vector ${\bf R}=R(\cos\phi,\sin\phi)$ which connects the two impurities, is determined by its length $R$ and angle $\phi$ with respect to the $x$-axis. 
Then, by Fourier expansion of the ${\bf k}$-dependent Green's function in terms of its angular harmonics as
\begin{eqnarray}
\check{G}_\omega({\bf k})  = \sum_m \check{g}_{m}(k,\omega) e^{im\theta_k},
\label{k-space-fourier}
\end{eqnarray}
and invoking the mathematical relation
\begin{eqnarray}
\int_{0}^{2\pi} \frac{d\theta_{k}}{2\pi}~ 
e^{i kR \cos (\theta_{k}-\phi) +im\theta_k}
= i^m J_m(kR) e^{in\phi}, 
\end{eqnarray}
in which $J_m(x)$ denote the Bessel functions of the first kind, we arrive at
\begin{eqnarray}
\check{G}_\omega({\bf R})&=&
\sum_m
 \check{\cal G}_m({R,\omega}) e^{i m\phi },
 \label{angular harmonics}\\
% \end{eqnarray}
%with the coefficients 
%\begin{eqnarray}
\check{\cal G}_m (R,\omega)&=&
\int\,\dfrac{d^2{\bf k}}{(2\pi)^{2}} \,  
i^mJ_m(kR) e^{-im\theta_{k}}  \: \check{G}_\omega({\bf k}). \label{coeffs-green} \end{eqnarray}
Above Eqs. show that the real-space Green's function 
$\check{G}_\omega({\bf R})$ can be expanded in terms of angular harmonics 
$e^{i m\phi }$ with coefficients $\check{\cal G}_m (R,\omega)$ which are respectively related to the coefficients $\check{g}_{m}(k,\omega)$
%=\int({d\theta_k}/2\pi)\, e^{-im\theta_{k}} \check{G}_\omega({\bf k})$ 
in the expansion (\ref{k-space-fourier}).
\par
Considering the perturbative form of the Green's function given by Eq. (\ref{green-0th-1st}), corresponding relations can be deduced for the expansion coefficients in Eqs. (\ref{k-space-fourier}) and (\ref{angular harmonics}) as
%$\check{g}_m(k,\omega)\approx\check{g}_m^{(0)}(k,\omega)+\check{g}_m^{(1)}(k,\omega)$
%and $\check{\cal G}_m(R,\omega)\approx\check{\cal G}_m^{(0)}(R,\omega)+\check{\cal G}_m^{(1)}(R,\omega)$, 
\begin{eqnarray}
\check{g}_m(k,\omega)&\approx& \check{g}_m^{(0)}(k,\omega)+\check{g}_m^{(1)}(k,\omega),\\
\check{\cal G}_m(R,\omega)&\approx&\check{\cal G}_m^{(0)}(R,\omega)+\check{\cal G}_m^{(1)}(R,\omega), 
\end{eqnarray}
respectively. Using the general relation (\ref{coeffs-green}) for the angular harmonic coefficients $\check{\cal G}_m$ in real space, we can immediately deduce
\begin{eqnarray}
\check{\cal G}_m^{(i)}(R,\omega)=
\int_0^{\infty}\,\dfrac{kdk }{2\pi} \,  
i^mJ_m(kR)  \: \check{g}_m^{(i)}(k,\omega). 
\label{coeffs-cal-g} \end{eqnarray}
for each perturbative term. It can be readily checked that the zeroth and first order terms of $\check{g}_{m}$ are given by
\begin{eqnarray}
\check{g}^{(0)}_{m}(k,\omega)&=&\int_0^{2\pi} \frac{d\theta_k}{2\pi}\, e^{-im\theta_{k}} \: \check{G}^{(0)}_\omega({\bf k}),\\
\check{g}^{(1)}_{m}(k,\omega)&=&\int_0^{2\pi} \frac{d\theta_k}{2\pi}\, e^{-im\theta_{k}}  \: \check{G}^{(0)}_\omega({\bf k})  \check{\cal H}^{(1)}  \check{G}^{(0)}_\omega({\bf k})\nonumber\\
&=&
\alpha_4 k \sum_{n}
\check{g}^{(0)}_{m-n} 
\big( 
\check{\cal T}_+ \:
\check{g}^{(0)}_{n+1}
+
\check{\cal T}_- \:
\check{g}^{(0)}_{n-1}
\big),\label{1st-order-calG}
\end{eqnarray}
where, to obtain the last line, the Fourier decomposition of Hamiltonian (\ref{asoc}) has been used:
\begin{eqnarray}
\check{\cal H}^{(1)}=\alpha_4 k (e^{i\theta_k}\check{\cal T}_- + e^{-i\theta_k}\check{\cal T}_+),~~\check{\cal T}_{\pm}=\frac{\check{\cal T}_x\pm i \check{\cal T}_y}{2}.~
\end{eqnarray}
It should be noted that the arguments of 
functions $\check{g}_{m}(k,\omega)$ inside the summation in Eq. (\ref{1st-order-calG}) has been dropped for the sake of compactness. 
As it is explained in the Appendix and can be understood from the explicit relation (\ref{green-2d}) 
for the zeroth order Green's function, its nominator consists of only $m=0,\pm2$ angular harmonics. Nevertheless, due to the presence of the denominator term ${\cal D}$ given by (\ref{d}), which can be decomposed as ${\cal D}={\cal D}_0-{\cal D}_1 \cos(4\theta_k)$ with 
${\cal D}_1\propto \alpha_3 (2\alpha_2+\alpha_3)$, all even angular harmonics of $\check{G}_{\omega}^{(0)}({\bf k})$ or equivalently $\check{G}_{\omega}^{(0)}({\bf R})$ are nonvanishing (More details can be found in the Appendix). Therefore, we can conclude that the angular-harmonics expansion coefficients $\check{g}_{m}^{(0)}(k,\omega)$ are nonzero only for even $m$'s. This is indeed an essential feature of half-Heusler TSM where the anisotropic $\alpha_3$ term in the Hamiltonian (\ref{luttinger-kohn}) is not negligible. Then, Eq. (\ref{1st-order-calG}) implies that  only $\check{g}^{(1)}_{m}(k,\omega)$ corresponding to odd $m$'s are not vanishing, a property which in fact originates from the $\theta_k$-dependence of $\check{H}^{(1)}$ with only $m=\pm 1$ harmonics. One can easily see that  due to the relation (\ref{coeffs-cal-g}), these properties of $\check{g}_{m}^{(i)}(k,\omega)$ ($i=0,1$) are inherited for real-space coefficients $\check{\cal G}_{m}^{(0)}(R,\omega)$ and $\check{\cal G}^{(1)}_{m}(R,\omega)$ which means they are respectively nonvanishing for even and odd $m$'s.
\par
In our perturbative scheme, the susceptibility up to the first order in the ASOC term, can be written as,
\begin{eqnarray}
&&{ \chi}_{ij}({\bf R}) \approx { \chi}_{ij}^{(0)}({\bf R})+{ \chi}_{ij}^{(1)}({\bf R}),%+{ \chi}_{ij}^{(1')}({\bf R}),
\end{eqnarray}
in which the explicit form of zeroth order term reads
\begin{eqnarray}
&&{ \chi}_{ij}^{(0)}({\bf R}) = 
\Im\int_{-\infty}^{\varepsilon_F} \frac{d\omega}{\pi}\, {\rm Tr}
\left[\check{\cal J}_i \check{G}^{(0)}_\omega({\bf R})  \check{\cal J}_j \check{G}^{(0)}_\omega(-{\bf R})\right] 
%\label{xi-perturb-1}
\nonumber\\
&&~~~=\Im \sum_{m,n} (-1)^n e^{im\phi} \int_{-\infty}^{\varepsilon_F} \frac{d\omega}{\pi}  
\, {\rm Tr} \left[ \check{\mathcal J}_{i}   \check{\mathcal G}^{(0)}_{m-n}     \check{\mathcal J}_{j}   \check{\mathcal G}^{(0)}_n     \right],~~~~
\label{xi-perturb-2}
\end{eqnarray}
where in the second line, the angular harmonics expansion (\ref{angular harmonics}) has been used. 
In a similar way, the explicit relations of the first order term ${\chi}_{ij}^{(1)}({\bf R})$ becomes
%by substituting the first and the second Green's functions in Eq. (\ref{xi-perturb-2}) with corresponding first order Green's functions $\check{G}_{\omega}^{(1)}$ and $\check{\cal G}^{(1)}_m$, respectively. So ${\chi}_{ij}^{(1)}({\bf R})$ is given by
\begin{eqnarray}
{ \chi}_{ij}^{(1)}({\bf R}) &=& 
\Im \sum_{m,n} (-1)^n e^{im\phi} \int_{-\infty}^{\varepsilon_F} \frac{d\omega}{\pi}  \nonumber\\
\, &\times& {\rm Tr} \left[ \check{\mathcal J}_{i}   \check{\mathcal G}^{(1)}_{m-n}     \check{\mathcal J}_{j}   \check{\mathcal G}^{(0)}_n +
\check{\mathcal J}_{i}   \check{\mathcal G}^{(0)}_{m-n}     \check{\mathcal J}_{j}   \check{\mathcal G}^{(1)}_n
\right],~~
\label{xi-perturb-1st}
\end{eqnarray}
consisting of two terms in which either the first or the second Green's functions appearing in the susceptibility relation is first order while another one is zeroth order.
Invoking the Eq. (\ref{coeffs-cal-g}), we obtain the following expressions for the zeroth- and first-order susceptibilities: 
\begin{widetext}
\begin{eqnarray}
\label{xi-perturb-zero}
{ \chi}_{ij}^{(0)}({\bf R}) &=& 
\Im \sum_{m,n} (-1)^n i^m e^{im\phi} \int_{-\infty}^{\varepsilon_F} \frac{d\omega}{\pi} \int \frac{k dk}{2\pi}\int \frac{k' dk'}{2\pi}
J_{m-n}(kR)J_{n}(k'R) \, {\rm Tr} \left[ \check{\mathcal J}_{i}   \check{g}^{(0)}_{m-n} (k,\omega)     \check{\mathcal J}_{j}   \check{g}^{(0)}_n  (k',\omega)    \right], \\
{ \chi}_{ij}^{(1)}({\bf R}) &=& 
\Im \sum_{m,n} (-1)^n i^m e^{im\phi} \int_{-\infty}^{\varepsilon_F} \frac{d\omega}{\pi} \int \frac{k dk}{2\pi}\int \frac{k' dk'}{2\pi}
J_{m-n}(kR)J_{n}(k'R) \nonumber \\
&& \qquad\qquad\qquad\qquad\qquad\qquad \quad \times \; {\rm Tr} \left[ \check{\mathcal J}_{i}   \check{g}^{(1)}_{m-n} (k,\omega)     \check{\mathcal J}_{j}   \check{g}^{(0)}_n  (k',\omega)  +\check{\mathcal J}_{i}   \check{g}^{(0)}_{m-n} (k,\omega)     \check{\mathcal J}_{j}   \check{g}^{(1)}_n  (k',\omega)    \right].
\label{xi-perturb-first}
\end{eqnarray}
\end{widetext}
We should remind that the explicit form of $\check{g}^{(0)}_{m} (k,\omega)$ are given by Eq. (\ref{explicit-g0}) in the Appendix and by inserting it in Eq. (\ref{1st-order-calG}), 
the first order terms $\check{g}^{(1)}_{m} (k,\omega)$ can be also explicitly obtained. Before going forward, it is worth to note that the presence of the factors $(-1)^n$ in above relations originates from the fact that the second Green's function in the susceptibility relation must be calculated at $-{\bf R}$ or equivalently $(R,\phi+\pi)$ in polar coordinates which immediately leads to an extra factor $e^{in\pi}=(-1)^n$ in the expansion (\ref{angular harmonics}) for $\check{G}^{(0)}_\omega(-{\bf R})$.
If we look back to the discussion after Eq. (\ref{1st-order-calG}), because of the fact that only even angular harmonics of $\check{G}^{(0)}_\omega$ are nonvanishing, then the $(-1)^n$ factor becomes irrelevant yielding that $\check{G}^{(0)}_\omega(-{\bf R})=\check{G}^{(0)}_\omega({\bf R})$.
However, following the same argument for $\check{G}^{(1)}_\omega(-{\bf R})$ which consists of only odd angular harmonics, we can easily check that $\check{G}^{(1)}_\omega(-{\bf R})=-\check{G}^{(1)}_\omega({\bf R})$. Subsequently, we can deduce different symmetry properties for the zeroth and first order susceptibilities as 
\begin{eqnarray}
\label{sym-1}
&&{\chi}_{ij}^{(0)}({\bf R})={ \chi}_{ij}^{(0)}(-{\bf R})={ \chi}_{ji}^{(0)}({\bf R}),\\
\label{sym-2}
&&{\chi}_{ij}^{(1)}({\bf R})=-{ \chi}_{ij}^{(1)}(-{\bf R})=-{\chi}_{ji}^{(1)}({\bf R}).
\end{eqnarray}
%The same properties as for ${\chi}_{ij}^{(1)}({\bf R})$ are true for ${\chi}_{ij}^{(1')}({\bf R})$.
Based on above symmetries, we can easily understand how bare LK Hamiltonian represented in ${\chi}_{ij}^{(0)}$ gives rise to a general Heisenberg-type RKKY interaction with a symmetric coupling matrix, while the first order corrections ${\chi}_{ij}^{(1)}({\bf R})$ due to the linear ASOC, lead to DM-type terms which are antisymmetric under the exchange of the two impurities \cite{dzyaloshinsky1958,moriya-prl-1960,moriya-pr-1960}.
\par
%The different components of the spin susceptibility and RKKY interaction strengths originating from bare KL Hamiltonian in the absence of ASOC, can be obtained by combining Eqs. (\ref{green-2d}), (\ref{coeffs-green}), and (\ref{xi-perturb-2}) and performing the numerical integration over the energy and momenta. Then, to find the first order corrections due to the ASOC, ${\chi}_{ij}^{(1)}({\bf R})$, we need to evaluate $\check{\cal G}^{(1)}_m$ numerically based on Eq. (\ref{1st-order-calG}). In the next section, we will present the results for all components of the RKKY coupling matrix obtained by the aforementioned procedure.
The different components of the spin susceptibility and RKKY interaction strengths originating from the LK Hamiltonian can be obtained by evaluating the integrals over the energy and momenta in Eq. (\ref{xi-perturb-zero}) numerically. In the same way and using Eq. (\ref{xi-perturb-first}) the first order correction ${\chi}_{ij}^{(1)}({\bf R})$ due to the ASOC term are calculated. For both cases, instead of performing the integration from $-\infty$, we use an energy cut-off $\Lambda=-2$ eV in consistency with the typical bandwidth of the realistic system based on the ab initio results \cite{felser2010natmat,hasan2010natmat,xiao2010halfheusler}. In the next section, the numerical results for all components of the RKKY coupling matrix obtained by the aforementioned procedure are presented.

%%%%%%%%%%%%%%%%%%%%%%%%%%%%%%%%%%%%%%%%%%%%%%%%%%%
%%%%%%%%%                              RESULTS SECTION                     %%%%%%%%%%%%%%%
%%%%%%%%%%%%%%%%%%%%%%%%%%%%%%%%%%%%%%%%%%%%%%%%%%%

\section{Numerical results and discussion}
\label{sec-res}
We saw in the previous section that because of relatively small effect of ASOC, it can be treated perturbatively, which enables us to separate
the spin susceptibility of the system to zeroth- and first-order parts with respect to ASOC term. So, we start by presenting the results for RKKY interaction mediated by LK 
part of the low-energy Hamiltonian as represented by the zeroth order susceptibility $\chi^{(0)}_{ij}$.
Then, we go through the first order contribution $\chi^{(1)}_{ij}$ which yield the DM-type RKKY coupling.  
\begin{figure*}[htp]
    \centering\includegraphics[width=.95 \textwidth]{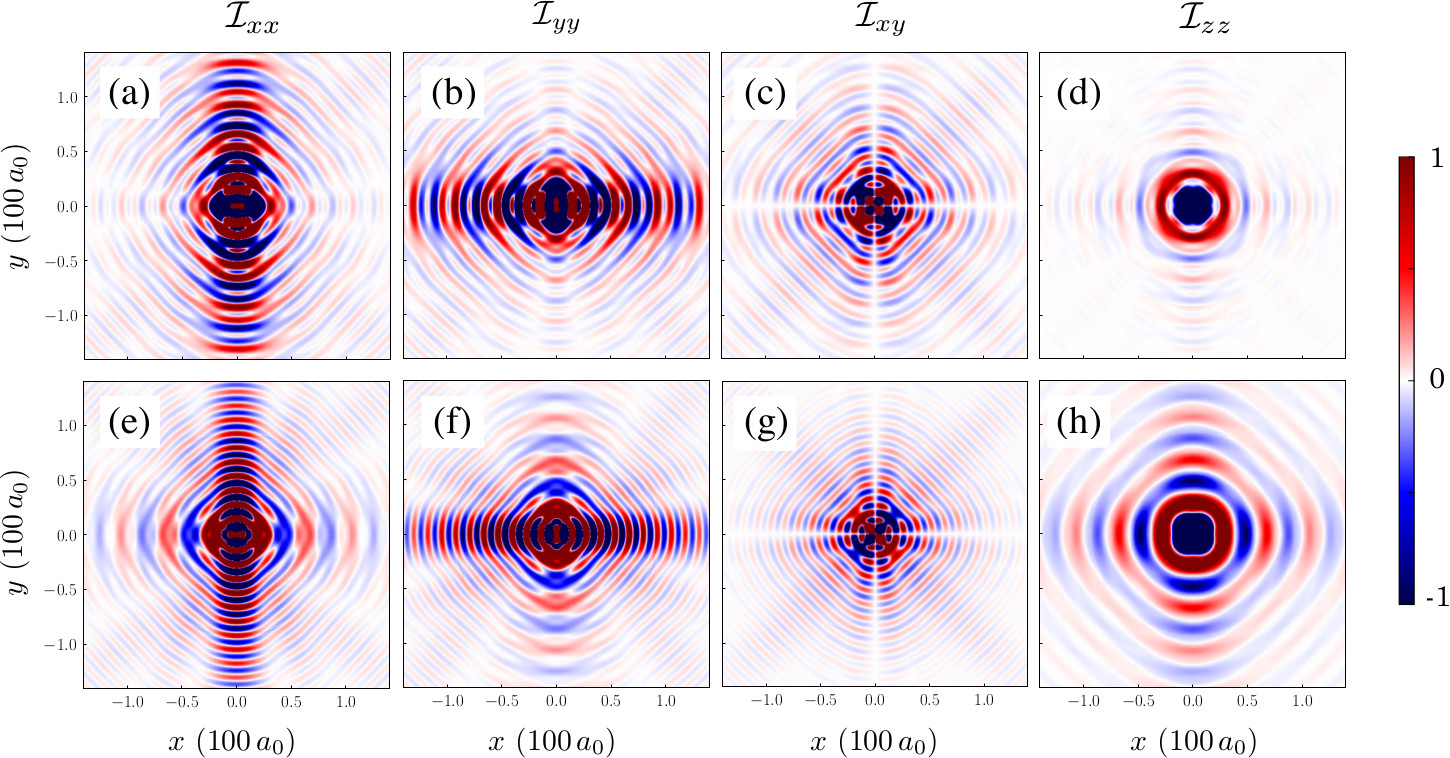}
    \caption{(Color online)
Spatial variation of symmetric components of the scaled RKKY interaction matrix ${\cal I}_{ij}$. As it has been discussed in the main text, among six possible symmetric components, ${\cal I}_{xz}$ and ${\cal I}_{yz}$ vanishes and only the diagonal terms as well as the off-diagonal term corresponding to in-plane spin orientation ${\cal I}_{xy}$ are finite. While the upper row (a-d) of subplots correspond to an electron-doped case ($\varepsilon_{F}=0.1$eV), the lower row (e-h) shows the corresponding results for hole doping with $\varepsilon_{F}=-0.1$eV. All RKKY components reveal three main features: strong anisotropy, tetragonal warping, and complex beating behavior in their oscillations. The qualitative difference between the electron- and hole-doped cases can be attributed to the electron-hole asymmetry already present in the band structure. Other parameters used for these plots are the same as those used to obtain Fig. \ref{fig0}. 
}\label{fig1}
\end{figure*}

\subsection{Generalized Heisenberg-type couplings}
As explained before, the symmetries of the LK Hamiltonian and the corresponding Green's function with only even angular harmonics,
lead to a symmetric zeroth-order susceptibility $\chi^{(0)}_{ij}({\bf R})$. Accordingly, the RKKY interaction given by Eq. (\ref{rkky}) takes the form of a generalized Heisenberg Hamiltonian
\begin{eqnarray}
{\cal H}^{(0)}_{\rm RKKY}\propto  \sum_{i,j}   {\sigma}_{1,i}  {\cal I}_{ij} ({\bf R}) {\sigma}_{2,j},\label{heisenberg}
\end{eqnarray}
in which the matrix elements ${\cal I}_{ij} \propto \chi^{(0)}_{ij}$ represent the dimensionless strengths of different components of RKKY interaction. To see why this Hamiltonian is of a generalized Heisenberg type, we note that the symmetry relation (\ref{sym-1}) implies that ${\cal I}_{ij}$ is also a symmetric matrix and therefore can be diagonalized.
Subsequently, in the absence of ASOC term, the energetically favored spin orientation of the two impurities will be collinear with an easy axis strongly depending on the relative position of the impurities. By direct inspection, we further realize that the off-diagonal terms of ${\cal I}_{ij} ({\bf R})$ in the unrotated (natural) frame before diagonalization, 
originate from the terms proportional to $\{{\cal J}_i ,{\cal J}_j\}$ ($i\neq j$) with coefficient $\alpha_2$ in the LK Hamiltonian.
So if we consider a bulk 3D TSM, we expect that all three symmetric off-diagonal terms in the RKKY Hamiltonian (\ref{heisenberg}) are present. But as a consequence of 2D geometry, only the in-plane off-diagonal component which is ${\cal I}_{xy}$ in our case does not vanish due to the presence of $\{{\cal J}_x ,{\cal J}_y\}$ term in the corresponding 2D form of LK Hamiltonian. Hence, in the quasi-2D structure considered here, a built-in anisotropy in the matrix structure of RKKY couplings exists yielding ${\cal I}_{xz}={\cal I}_{yz}=0$. Furthermore, on the same ground and because of the reduced dimensionality again, the diagonal terms corresponding to the in-plane spin components ${\cal I}_{xx}$ and ${\cal I}_{yy}$ behave quite differently than the out-of-plane component ${\cal I}_{zz}$. This means that even the diagonal terms of RKKY coupling matrix alone give rise to an anisotropic Heisenberg model which can be decomposed to an Ising-type term along $z$ direction and an isotropic Heisenberg Hamiltonian. \par
Now after overviewing some general features of the RKKY Hamiltonian (\ref{heisenberg}), we present the spatial variation of its different components obtained by the method developed in Sec. \ref{sec-2}. By direct evaluation, we find the following Fourier expansion for the nonvanishing components in the 2D case:
\begin{eqnarray}
&&{\cal I}_{xx}({\bf R})={\cal I}_{xx,0}(R)+{\cal I}_{xx,2}(R)\cos(2\phi) +\cdots ,\\
&&{\cal I}_{yy}({\bf R})={\cal I}_{xx,0}(R)-{\cal I}_{xx,2}(R)\cos(2\phi) +\cdots ,\\
&&{\cal I}_{xy}({\bf R})={\cal I}_{xy,2}(R)\sin(2\phi) +\cdots , \label{IXY} \\
&&{\cal I}_{zz}({\bf R})={\cal I}_{zz,0}(R)+{\cal I}_{zz,4}(R)\cos(4\phi) +\cdots .\label{IZZ}
\end{eqnarray}
For the sake of compactness of the relations, only the lowest nonvanishing angular-harmonic terms are explicitly written without presenting too lengthy forms of the distance dependent coefficients ${\cal I}_{ij,m}({R})$. Instead, the full numerical results for the spatial dependence of the RKKY couplings are shown in Fig. \ref{fig1} which reveal their interesting properties. We should mention that higher harmonics [$\cos(2m\phi)$ or $\sin(2m\phi)$] are also present and as it is clear from the contour plots their contributions are not negligible. In fact, all nonvanishing four components are strongly anisotropic and significantly vary with the angle $\phi$ of the impurities relative locations. In other words, while keeping the distance $R$ between the impurities fixed and rotating the second impurity around the first one, the RKKY couplings drastically change and even undergo sign changes.
The anisotropic features in the spatial variation of the RKKY interaction indeed originates from the strong anisotropy of the band structure and $\theta_k$-dependence of the Green's function (\ref{green-2d}). Furthermore, all components clearly demonstrate a tetragonally warped spatial dependence dictated by similar property of the Fermi lines as seen in Fig. \ref{fig0}. 
\par
As another general feature of all components of the RKKY coupling matrix, a beating pattern with distinct periodicities is recognizable. The appearance of such beating pattern relies on the tetragonally warped Fermi lines which give rise to a range of Fermi wavelengths, a property far different from conventional electronic systems with almost circular Fermi lines and a single Fermi wavelength. Typically, the beating pattern in RKKY oscillations can be found in electronic systems with two or more Fermi surfaces, including the well-known example of 2D Rashba gases and recently found 3D Rashba systems \cite{loss2013,rkky-shiba-3d}. In contrast, we have essentially a single Fermi line (more precisely there are two degenerate ones) before the inclusion of ASOC effects. Hence, because of the strong warping of the Fermi line, the magnitude of the Fermi wave vector does depend on the direction of momentum which eventually leads to a complex beating pattern in the 2D map of RKKY oscillations. Interestingly, the periodicity of oscillations with the distance $R$ varies by changing the angle $\phi$ due to the strong warping feature. 
All these features are almost unique to the TSMs and are not seen in other electronic systems specially those with SOCs \cite{bruno-prb-2004,chao2007,loss2008,egger2009,jiaji2010,loss2013,rkky-shiba-3d,tse-bulk-rashba,saha2019}. 
To illustrate the role of chemical doping, Figs. \ref{fig1}(a)-(d) and \ref{fig1}(e)-(h) show the results for electron- and hole-doped cases with $\varepsilon_F=\pm 0.1 eV$, respectively. 
By pairwise comparison of the plots, we see each ${\cal I}_{ij}$ possesses almost similar qualitative spatial variations for electron- and hole-doped cases.
This observation is consistent with the same property in the band structure of the system in which electron and hole dispersions only differ in their effective masses. Then, the only effect of different effective masses is reflected in the different periodicities of RKKY oscillations.
\par
Now, let us separately elucidate the properties of each component ${\cal I}_{ij}$ based on the results shown in Fig. \ref{fig1}. First, we see that the diagonal elements, corresponding to the interaction of the in-plane components of the two spins, have exactly the same variation with ${\bf R}$ except for a $\pi/2$ rotation, a property which has its root in the symmetries of the band structure and the unperturbed Green's function (\ref{green-2d}). Intriguingly, the dimensionless couplings ${\cal I}_{xx}$ and ${\cal I}_{yy}$ are stronger around vertical and horizontal directions which correspond to positioning of two impurities along $y$ and $x$ directions, respectively. This, in fact, manifests the dominance of first nonvanishing harmonic $\cos(2\phi)$, although the higher harmonics are also not negligible at all. So comparing Figs. \ref{fig1}(a) and (b) or Figs. \ref{fig1}(e) and (f), we see an strong anisotropy between two in-plane components ${\cal I}_{xx}$ and ${\cal I}_{yy}$ but in a way their dependence on $\phi$ is actually related to each other by a $\pi/2$ rotation. The in-plane anisotropy which can be regarded as Ising-type term along $x$ or $y$ direction, is very special to our system and SSOC terms of the LK Hamiltonian. In the 3D limit of our system, one can expect similar anisotropy between all three diagonal couplings ${\cal I}_{ii}$. This is in contrast to other spin-orbit coupled systems including 2D and 3D Rashba materials in which only perpendicular Ising-type term along $z$ direction has been found \cite{bruno-prb-2004,rkky-shiba-3d,tse-bulk-rashba}. Now we arrive in the off-diagonal coupling ${\cal I}_{xy}$
which exposes a fan-like shape in its spatial oscillations originating from the $\sin(2\phi)$ and higher terms in Eq. (\ref{IXY}). As mentioned before, the nonvanishing off-diagonal RKKY component comes from the SSOC term proportional to $\{{\cal J}_x ,{\cal J}_y\}$ in the LK Hamiltonian and hence the special $\phi$-dependence of ${\cal I}_{xy}$ can be attributed to the momentum dependence of the corresponding term in the Green's function (\ref{green-2d}) with the explicit form $k_x k_y\equiv k^2 \sin(2\theta_k)/2$. Finally, as we see from Figs. \ref{fig1} (d) and (h), the perpendicular component ${\cal I}_{zz}$ is far different from the other two diagonal terms, in accordance with the 2D nature of the system as we discussed before. Looking to the Fourier expansion (\ref{IZZ}) for ${\cal I}_{zz}$ we see that it includes only the harmonics of the form $\cos(4n\phi)$ which yield a $\pi/2$-periodic dependence on the angle rather than $\pi$-periodicity observed for the in-plane components of the RKKY interaction matrix. 
\subsection{Dzyaloshinskii-Moriya-type couplings}
\begin{figure}[t!]
    \centering\includegraphics[width=.9 \linewidth]{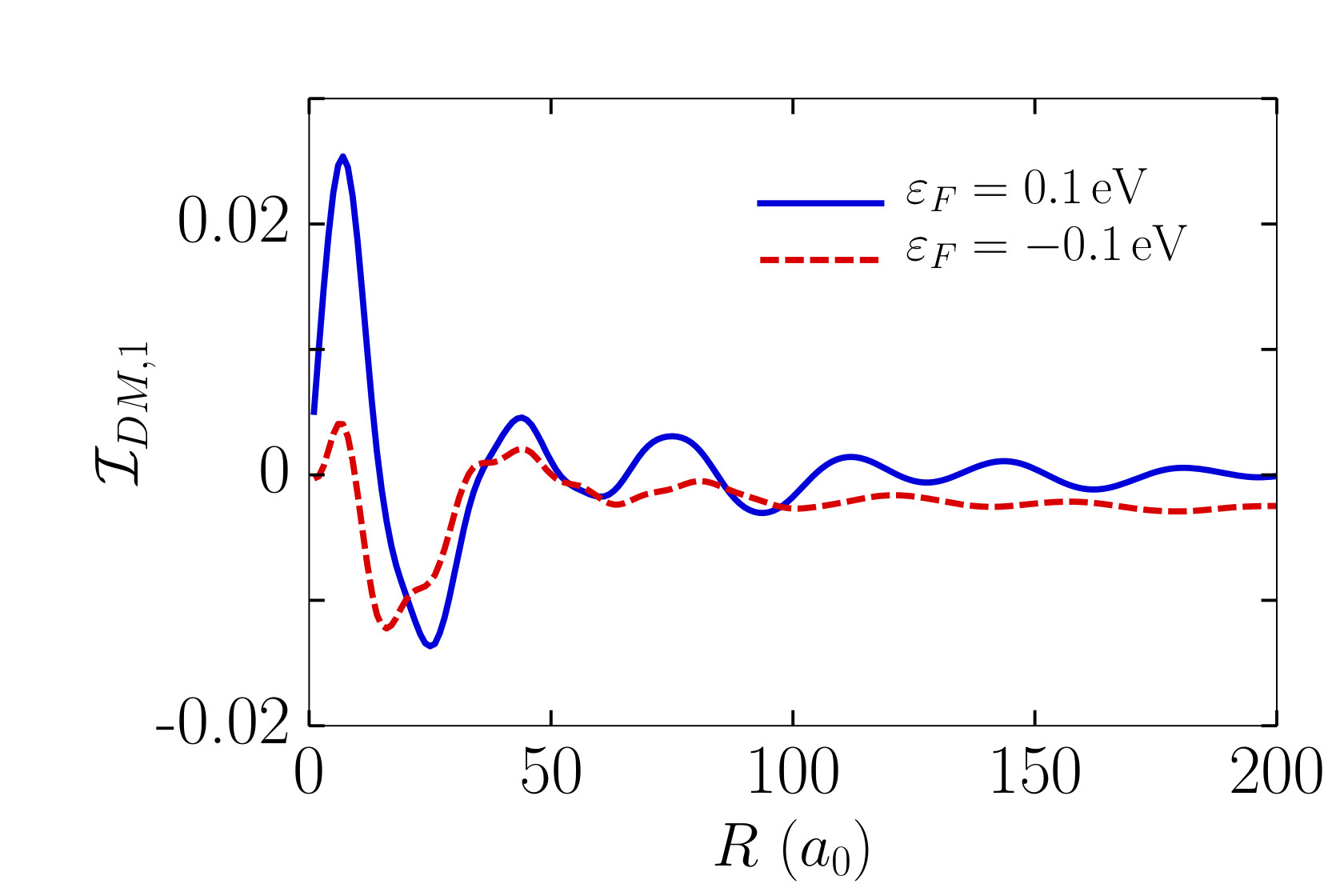}
    \caption{(Color online)
    Spatial variation of the first harming coefficient of DM-type couplings, ${\cal I}_{{\rm DM},1}$ for an electron- (solid blue line) and hole-doped (dashed red line) cases. The qualitative behavior in electron and hole dopings is similar and we only see slight differences due to the electron-hole asymmetry in the same way as Heisenberg-type terms. Here we consider $\alpha_4 \sim0.1\,(a_{0}/\pi)$eV and other parameters are the same as those used to obtain Fig. \ref{fig0}.
}
\label{fig3}
\end{figure}
As we have already pointed in Sec. \ref{sec-2}, by taking the ASOC term into account and considering only first order corrections ${\chi}^{(1)}_{ij}$ to the spin susceptibility, antisymmetric matrix elements in the RKKY coupling matrix show up. These new terms are represented by dimensionless quantities ${\cal I}'_{ij} \propto {\chi}^{(1)}_{ij}$ with the same prefactor used in relating symmetric components ${\cal I}_{ij}$ to the zeroth order susceptibilities ${\chi}^{(0)}_{ij}$.
So, we can write down the corresponding contribution to the RKKY Hamiltonian in terms of a so-called DM vector ${\mathbfcal I}_{{\rm DM}}$
with components ${\cal I}_{{\rm DM},i}=\varepsilon_{ijk}{{\cal I}'_{jk}}/2$, as 
\begin{eqnarray}
{\cal H}^{(1)}_{\rm RKKY}\propto  \sum_{i,j} {\mathbfcal I}_{{\rm DM}} 
\cdot
 \left( {\bm \sigma}_{1}\times {\bm \sigma}_{2} \right).
\end{eqnarray}
By direct evaluation of the first order correction to the Green's function and the susceptibility given by Eqs. (\ref{1st-order-calG})
and (\ref{xi-perturb-first}), respectively, we find that only ${\cal I}'_{xz}$ and ${\cal I}'_{yz}$ are not vanishing. Subsequently, we obtain the following two components of the DM vector
\begin{eqnarray}
&&{\cal I}_{{\rm DM},x}({\bf R})={\cal I}_{{\rm DM},1}(R)\cos\phi +\cdots ,\\
&&{\cal I}_{{\rm DM},y}({\bf R})={\cal I}_{{\rm DM},1}(R)\sin\phi +\cdots ,
\end{eqnarray}
originating respectively from the ASOC terms $k_x {\cal T}_x$ and $k_y {\cal T}_y$. As discussed before, the first order corrections to the spin susceptibility and the resulting DM-type RKKY terms include all odd harmonics in their expansion in terms of angular harmonics. Therefore, similar to the Heisenberg-type terms, the DM terms possess features like strong anisotropy, tetragonal warping, and beating pattern. In particular, the absence of third component of DM-type couplings is a consequence of 2D geometry of the system. But if we consider a bulk 3D system, all three DM components should exist.
\par
Figure \ref{fig3} shows the variation of ${\cal I}_{{\rm DM},1}$ with the distance $R$ of the two impurities
for two values of chemical potential $\varepsilon_F=\pm0.1$ eV.  As expected, we see that the relative strength of DM terms is much smaller than the Heisenberg ones (about two orders of magnitude at the chemical potential of $|\varepsilon_F|=0.1$ eV). Besides, similar to the other components and as a common key feature of the RKKY coupling, the DM terms also have oscillatory decaying dependence on the distance with almost the same qualitative behavior for both electron- and hole-doped cases. However, due to the anisotropic and tetragonally-warped isoenergy lines, the oscillation pattern is not as regular as standard RKKY interaction in conventional electron gases with a unique Fermi wave vector $k_F$.
\par
It must be mentioned that in spite of their relatively smaller strengths, the two nonvanishing DM terms or equivalently ${\cal I}'_{xz}$ and ${\cal I}'_{yz}$ should not be ignored. This is essentially due to the fact that their corresponding terms in the symmetric (Heisenberg-type) part of the RKKY Hamiltonian are identically zero. Considering the full matrix of RKKY couplings by summing the zeroth- and first-order contributions as ${\cal I}^{\rm full}_{ij}={\cal I}_{ij}+{\cal I}'_{ij}$, we immediately see that among the off-diagonal terms, only ${\cal I}^{\rm full}_{xy}$ is symmetric and the other two terms are fully antisymmetric and represent DM-type interaction. This property of RKKY matrix originates from the 2D geometry of the system considered here, while in 3D, all off-diagonal elements have nonvanishing symmetric and antisymmetric contributions with the dominance of symmetric parts. It is worth to remind that the key property of DM-type interactions is that they lead to twisted spin-spin interaction and yield a noncollinear spin configuration \cite{bruno-prb-2004}. Hence, we should expect that at the vicinity of certain relative positions of the impurities where all Heisenberg terms become small, the DM terms come into play and induce significant twist in the spin orientation of the two impurities. 

\begin{figure}[t!]
    \centering\includegraphics[width=.99 \linewidth]{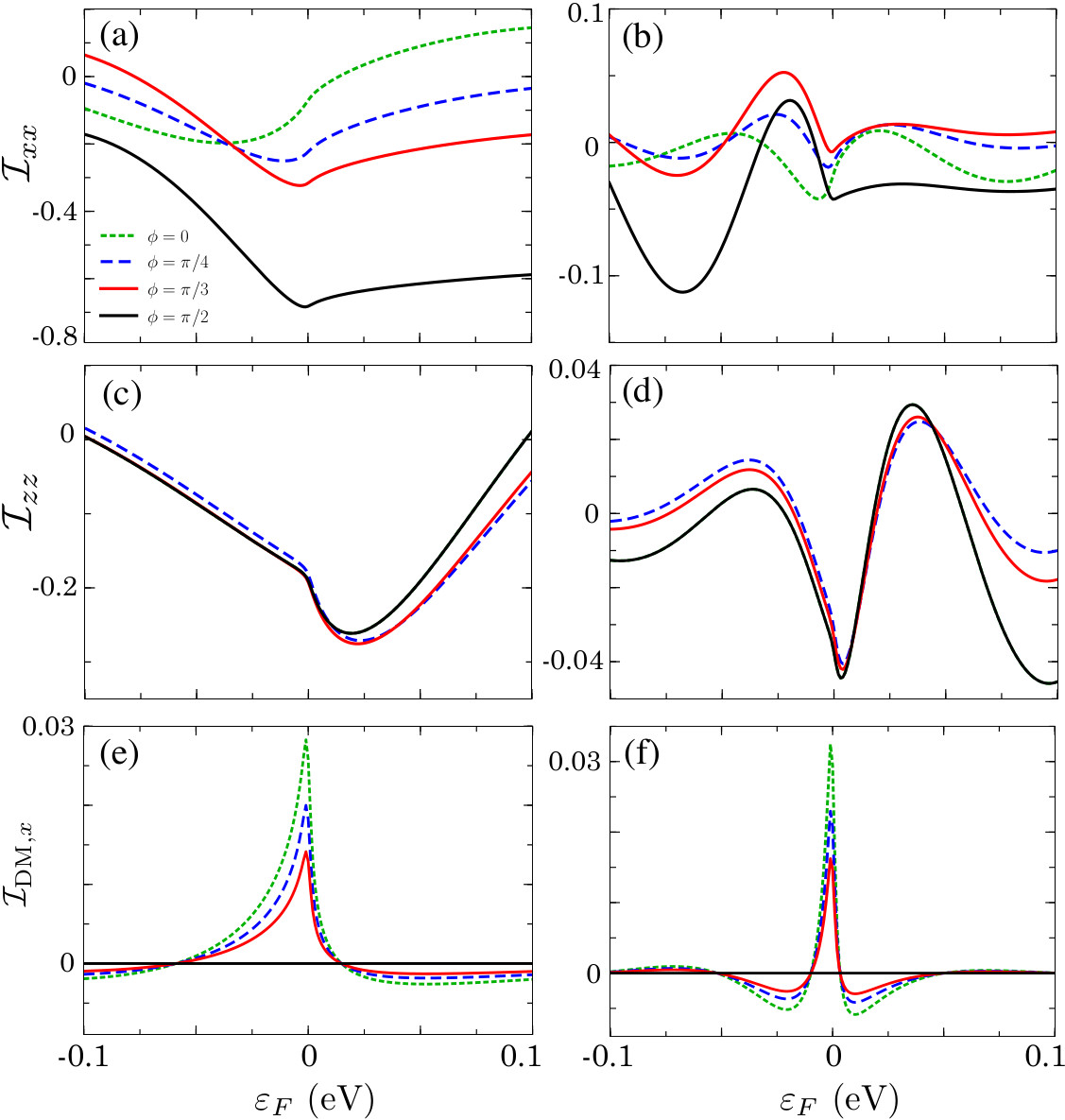}
    \caption{(Color online) 
    The variation of RKKY couplings with the Fermi energy at a constant distance $R$ and for different angles $\phi$ representing the relative position of the impurities. 
    Plots in the left and right sides correspond to the distances $R=20 a_{0}$ and $R=50 a_{0}$. We see that at the larger distance, while the strength of different components decrease, the rate of variations with the Fermi energy increases such that the oscillatory dependence on $\varepsilon_F$ is clearly seen in the results for $R=50 a_{0}$ (For the smaller distance, the oscillations are apparent for a wider energy range).
The in-plane diagonal coupling strength ${\cal I}_{xx}$ [shown in (a) and (b)] shows more significant dependence on the angle $\phi$ rather than perpendicular component ${\cal I}_{zz}$ [shown in (c) and (d)] and DM term ${\cal I}_{{\rm DM},x}$ [shown in (e) and (f)]. Finally as a general feature in all results, almost abrupt changes are observed around the band touching point where the Fermi momentum vanishes.   
All parameters used here are the same as those used to obtain Figs. \ref{fig0} and \ref{fig3}. 
    }\label{fig4}
\end{figure}

\subsection{Dependence on the Fermi energy}
So far, we have concentrated on two certain values of chemical potential $\varepsilon_F=\pm 0.1$ eV and it has been already shown that the spatial variations of all different terms of RKKY couplings are qualitatively the same for electron and hole dopings. To better elucidate the dependence of different RKKY terms on the Fermi energy, we show the evolution of ${\cal I}_{xx}$, ${\cal I}_{zz}$, and ${\cal I}_{{\rm DM},x}$ with $\varepsilon_F$ in Fig. \ref{fig4}, at two different distances $R$ and for various angles $\phi$. The results reveal a complex and irregular energy dependence for the range of energies considered and especially around the band touching point ($\varepsilon_F=0$). 
Particularly, at the larger distance $R=50 a_0$, an oscillatory decaying dependence on the Fermi energy can be recognized when we go away from $\varepsilon_F=0$ to either electron or hole dopings. This property besides the variation with the angle $\phi$ complements the similar qualitative picture for the spatial variations at constant energies shown in Fig. \ref{fig1}. Similarities in variations with either distance $R$ or energy $\varepsilon_F$ are among well-known aspects of the RKKY coupling and related phenomena like Friedel oscillations, which rely on the singular behavior of response functions for momentum transfers of two times the Fermi momentum ($\Delta k=2k_F$) \cite{solyom2008fundamentals}. Nevertheless, unlike the simple electronic systems with isotropic dispersion relations where RKKY interaction is a function of $\varepsilon_F R/\hbar v_F$, a unique Fermi momentum does not exist here. Therefore, the position and energy dependences cannot be expressed on the same footing and in terms of a single dimensionless parameter. 

\subsection{Comparison to other systems}
We have mentioned before that the RKKY coupling between magnetic impurities in half-Heusler TSMs which host effective spin-$3/2$ quasiparticles, has special signatures beyond the typical properties of the RKKY coupling in other electronic systems. Now, we try to elucidate these differences, particularly compared to the systems with effective spin-$1/2$ descriptions and at the presence of SOC terms.
In the most well-known example of 2D Rashba gas, the presence of SOC leads to DM-type couplings, while here, the ASOC term is responsible for these terms in the RKKY coupling. 
Nevertheless, the half-Heusler TSMs as described by LK Hamiltonian possess much stronger SSOC terms which are quadratic in both momentum and spin-$3/2$ operators. 
This symmetric SOC term $({\bf k}\cdot \check{\mathbfcal J})^2$ leads to the symmetric off-diagonal components, namely ${\cal I}_{xy}$ in our 2D case. We may remind while DM terms favor noncollinear 
spin alignments, the presence of symmetric off-diagonal Heisenberg terms give rise to an inclination of both spins with respect to the lattice of TSM or equivalently the natural $x-y$ coordinates. Therefore, due to the presence of off-diagonal RKKY component which is also strongly angle-dependent, the energetically favored spin direction of two impurities varies a lot with the angle $\phi$. This means that not only the strength of the couplings between the spins but also their preferred direction changes by rotating the impurities around each other. These remarkable features related to the ${\cal I}_{xy}$ RKKY term, are totally absent in spin-$1/2$ systems, since they originate from the SSOC which can exist only when the quasiparticles have higher spin as discussed in Sec. \ref{sec-2}. 
\par
Regarding the LK Hamiltonian, a few works have already studied RKKY coupling but within the isotropic model with $\alpha_3=\alpha_4=0$ which is typically valid for $p$-doped zinc blende semiconductors \cite{kernreiter13,verma19}. So they lack all sort of anisotropic features found here for the RKKY coupling of magnetic impurities in half-Heusler TSMs.
It should be mentioned that the key property of RKKY coupling in TSMs is that the anisotropic attributes are twofold in nature. In fact, not only each RKKY components have 
strong dependence on the relative angle $\phi$ of the impurities positions, but the matrix structure of the interaction also reveals anisotropies due to either the differences between the diagonal elements ${\cal I}_{ii}\neq {\cal I}_{jj}$ ($i\neq j$) or the presence of off-diagonal terms. We know that mere anisotropy in the spatial dependence of each RKKY component is not very specific to our system and it can be essentially found in other systems with anisotropic band structure \cite{Akbari-2013,asmar-2017}. Nevertheless, the combination of peculiar spatial dependence 
with the anisotropies in the matrix structure makes the RKKY coupling in TSMs far different from previously explored systems. For the 2D structures of TSM considered here, the presence of symmetric and antisymmetric SOCs which respectively yield ${\cal I}_{xy}$ and ${\mathbfcal I}_{{\rm DM}}$ is indispensable to see the aforementioned characteristics of the RKKY interaction.

\section{Conclusions}\label{conc}
We have investigated the indirect exchange coupling coined as RKKY interaction between magnetic impurities mediated by $j=3/2$ quasiparticles in half-Heusler topological semimetals confined in a two-dimensional geometry. To tackle this problem in which the dispersion relation and the bare Green's function are very anisotropic, we have developed a method to obtain RKKY couplings strengths in a series over angular harmonics $e^{im\phi}$ ($\phi$ is the polar angle of the 2D vector ${\bf R}$ connecting the two impurities). Then, as a general feature, it has been found that all RKKY couplings are strongly anisotropic and drastically change by rotating an impurity around another. Also, corresponding to the tetragonal warping of the isoenergy lines in the band structure, the dependence of RKKY couplings on ${\bf R}$ have the same behavior on top of the oscillatory decaying variations. Such features are very special to the half-Heusler semimetals and do not appear in the conventional electronic systems with almost isotropic band dispersion. 
Moreover, the matrix structure of the RKKY coupling is also quite rich possessing unequal diagonal components as well as various symmetric and antisymmetric off-diagonal terms. In particular, we have shown that the Kohn-Luttinger part of the effective Hamiltonian gives rise to a generalized Heisenberg coupling with anisotropic matrix structure due to the reduced dimensionality of the system and existence of symmetric spin-orbit coupling terms in close connection with the spin-$3/2$ nature of the quasiparticles. Then by considering the effect of antisymmetric linear spin-orbit coupling term which is also present in half-Heusler semimetals, Dzyaloshinskii-Moriya type RKKY couplings have been obtained.

Our findings which can be tested using spin-polarized scanning tunneling spectroscopy techniques, also implies interesting features for the magnetic ordering of dilute local moments in topological semimetals. Particularly, the anisotropy of the RKKY coupling and the presence of spin-twisting terms suggest noncollinear spin ordering between the magnetic impurities. Consequently, provided by the existence of a few superconducting half-Heusler materials,
the twisted spin alignment mediated by RKKY coupling may be promising to realize Majorana fermions in half-Heusler superconductors possessing spin-$3/2$ low-energy excitations.
In this regard, further investigations are demanded to better understand the impurity-related properties of the half-Heusler topological semimetals, particularly in connection with their magnetic and superconducting characters. 

\section{Acknowledgements}
The authors thank the International Center for Theoretical Physics (ICTP) in Trieste for the hospitality and support during a visit when part of this work was done. A.G.M. acknowledges financial support from Iran Science Elites Federation under Grant No. 11/66332. 

\appendix

\section{Angular harmonics expansion of $G_\omega^{(0)}({\bf k})$}
In this part, we derive the explicit form of the ${\bf k}$-space Green's function $G_\omega^{(0)}({\bf k})$ in terms of its angular harmonics $e^{im\theta_k}$.
We first note that only two terms 
$\alpha_2 ({\bf k}\cdot \check{\mathbfcal J})^2$ and $\alpha_3 \sum_{i=1}^{2}k_i^2 \check{\cal J}_i^2$ 
in the 
numerator of the Green's function given by Eq. (\ref{green-2d}) depend on the angle of momentum $\theta_k$. So by substituting $k_x=k \cos\theta_k$ and $k_y=k \sin\theta_k$ inside these terms, they can be written as 
\begin{eqnarray}
&&\alpha_2 k^2 (e^{i\theta_k} \check{\cal J}_-+e^{-i\theta_k} \check{\cal J}_+)^2,\\
&&
\alpha_3 k^2 (J_x^2 \cos^2\theta_k+J_y^2 \sin^2\theta_k),
\end{eqnarray}
respectively, where $\check{\cal J}_\pm=(\check{\cal J}_x\pm i\check{\cal J}_y)/2$. Both of these expressions can be re-written in terms of only three angular harmonics $e^{im\theta_k}$ with $m=0,\pm2$, which implies the same property for the whole numerator of the expression (\ref{green-2d}) for 
$G_\omega^{(0)}({\bf k})$. Therefore after some algebra we find,
\begin{eqnarray}
\check{G}^{(0)}_\omega({\bf k})=\dfrac{\check{\cal C}_0(k,\omega)+
e^{2i\theta_{k}}\:
\check{\cal C}_{+}(k)+e^{-2i\theta_{k}}\:\check{\cal C}_{-}(k) }{\cal D},~~
\end{eqnarray}
in which
\begin{eqnarray}
\check{\cal C}_0(k,\omega)&=&
\big[
\omega- \frac{5}{2}\beta-(\alpha_{1}+\frac{5}{8} \alpha_2 +\frac{5}{8} \alpha_3 ) k^{2} 
\big] \check{\mathbbm{1}}\nonumber\\
&+&
 \big[
\beta-(\frac{ \alpha_2+ \alpha_3}{2}) k^2
\big]\check{\cal J}_{z}^2,
\label{g-2d-harmonic-0} 
\\
\check{\cal C}_+(k,\omega)&=&
 (\alpha_2+ \alpha_3)\check{\cal J}_{-}^2+i\frac{\alpha_3}{4} 
 \{ \check{\cal J}_{x}, \check{\cal J}_{y}\},
 \label{g-2d-harmonic-p} 
 \\
\check{\cal C}_-(k,\omega)&=&
 (\alpha_2+ \alpha_3)\check{\cal J}_{+}^2-i\frac{\alpha_3}{4} 
 \{ \check{\cal J}_{x}, \check{\cal J}_{y}\}.
\label{g-2d-harmonic-m} 
\end{eqnarray}
Accordingly, we can express the denominator which has been given by (\ref{d}) in terms of three angular harmonics corresponding to $m=0,\pm4$ as    
${\cal D}= {\cal D}_0(k,\omega)-{\cal D}_1(k) \cos(4\theta_k)$ 
with
\begin{eqnarray}
{\cal D}_0(k,\omega)&=& [\omega-\epsilon_{k,+}][\omega-\epsilon_{k,-}],\\
\epsilon_{k,\pm}&=&\frac{5 \beta }{4}+k^2 \big[ 
\alpha_1+\frac{5}{4}(\alpha_2+\alpha_3)\big]
\pm
\sqrt{\Delta_k},~~\\
\Delta_k&=&
[\beta-\frac{k^2}{2}(\alpha_2+\alpha_3)]^2
\nonumber\\
&+&\frac{3}{4}k^4(\alpha_2^2+\frac{\alpha_3^2}{2}+\alpha_2\alpha_3),\\
{\cal D}_1(k)&=&
\frac{3}{8}  \alpha_3 (2\alpha_2+\alpha_3)k^4.
\end{eqnarray}
In order to find the corresponding expansion coefficients $\check{g}^{(0)}_m(k,\omega)$, we first expand the prefactor $1/{\cal D}$ in terms of $e^{im\theta_k}$ as the following:
\begin{eqnarray}
\frac{1}{\cal D}&=&\frac{1}{{\cal D}_0}\sum_{n=0}^{\infty}\sum_{m=-n}^{n}  \binom{n}{\frac{n+m}{2}} \big( \frac{{\cal D}_1}{{\cal D}_0} \big)^{n} e^{4im\theta_k}\nonumber\\
\label{binomial}
&=&\sum_{m=-\infty}^{\infty} e^{4im\theta_k} \sum_{n=|m|}^{\infty}  \binom{n}{\frac{n+m}{2}}  \frac{ {\cal D}_1^n}{ {\cal D}_0^{n+1}}.
\end{eqnarray}
The second line of above equation has been obtained by interchanging the order of two summation variables. Using this expression, we find the nonvanishing expansion coefficients of the Green's function for any integer $m$ as
\begin{eqnarray}
\check{g}^{(0)}_{4m}(k,\omega) &=&
\sum_{n=|m|}^{\infty}  \binom{n}{\frac{n+m}{2}}  \frac{ {\cal D}_1^n}{ {\cal D}_0^{n+1}} \: \check{\cal C}_0(k,\omega),\nonumber \\
\check{g}^{(0)}_{4m+2}(k,\omega) &=&
\sum_{n=|m|}^{\infty}  \binom{n}{\frac{n+m}{2}}  \frac{ {\cal D}_1^n}{ {\cal D}_0^{n+1}} \: \check{\cal C}_+(k,\omega)
\label{explicit-g0}
\\
&+&
\sum_{n=|m+1|}^{\infty}  \binom{n}{\frac{n+m+1}{2}}  \frac{ {\cal D}_1^n}{ {\cal D}_0^{n+1}} \: \check{\cal C}_-(k,\omega),\nonumber 
\end{eqnarray}
in which
\begin{eqnarray}
\binom{n}{m}=\frac{n!}{(n-m)!m!},
\end{eqnarray}
denotes the binomial coefficient.

\section*{References}
\bibliography{tsmrkky.bib}

\begin{thebibliography}{70}%
\makeatletter
\providecommand \@ifxundefined [1]{%
 \@ifx{#1\undefined}
}%
\providecommand \@ifnum [1]{%
 \ifnum #1\expandafter \@firstoftwo
 \else \expandafter \@secondoftwo
 \fi
}%
\providecommand \@ifx [1]{%
 \ifx #1\expandafter \@firstoftwo
 \else \expandafter \@secondoftwo
 \fi
}%
\providecommand \natexlab [1]{#1}%
\providecommand \enquote  [1]{``#1''}%
\providecommand \bibnamefont  [1]{#1}%
\providecommand \bibfnamefont [1]{#1}%
\providecommand \citenamefont [1]{#1}%
\providecommand \href@noop [0]{\@secondoftwo}%
\providecommand \href [0]{\begingroup \@sanitize@url \@href}%
\providecommand \@href[1]{\@@startlink{#1}\@@href}%
\providecommand \@@href[1]{\endgroup#1\@@endlink}%
\providecommand \@sanitize@url [0]{\catcode `\\12\catcode `\$12\catcode
  `\&12\catcode `\#12\catcode `\^12\catcode `\_12\catcode `\%12\relax}%
\providecommand \@@startlink[1]{}%
\providecommand \@@endlink[0]{}%
\providecommand \url  [0]{\begingroup\@sanitize@url \@url }%
\providecommand \@url [1]{\endgroup\@href {#1}{\urlprefix }}%
\providecommand \urlprefix  [0]{URL }%
\providecommand \Eprint [0]{\href }%
\providecommand \doibase [0]{http://dx.doi.org/}%
\providecommand \selectlanguage [0]{\@gobble}%
\providecommand \bibinfo  [0]{\@secondoftwo}%
\providecommand \bibfield  [0]{\@secondoftwo}%
\providecommand \translation [1]{[#1]}%
\providecommand \BibitemOpen [0]{}%
\providecommand \bibitemStop [0]{}%
\providecommand \bibitemNoStop [0]{.\EOS\space}%
\providecommand \EOS [0]{\spacefactor3000\relax}%
\providecommand \BibitemShut  [1]{\csname bibitem#1\endcsname}%
\let\auto@bib@innerbib\@empty
%</preamble>
\bibitem [{\citenamefont {Haldane}(2017)}]{haldane-nobel}%
  \BibitemOpen
  \bibfield  {author} {\bibinfo {author} {\bibfnamefont {F.~D.~M.}\
  \bibnamefont {Haldane}},\ }\href {\doibase 10.1103/RevModPhys.89.040502}
  {\bibfield  {journal} {\bibinfo  {journal} {Rev. Mod. Phys.}\ }\textbf
  {\bibinfo {volume} {89}},\ \bibinfo {pages} {040502} (\bibinfo {year}
  {2017})}\BibitemShut {NoStop}%
\bibitem [{\citenamefont {Hasan}\ and\ \citenamefont {Kane}(2010)}]{kane-rmp}%
  \BibitemOpen
  \bibfield  {author} {\bibinfo {author} {\bibfnamefont {M.~Z.}\ \bibnamefont
  {Hasan}}\ and\ \bibinfo {author} {\bibfnamefont {C.~L.}\ \bibnamefont
  {Kane}},\ }\href {\doibase 10.1103/RevModPhys.82.3045} {\bibfield  {journal}
  {\bibinfo  {journal} {Rev. Mod. Phys.}\ }\textbf {\bibinfo {volume} {82}},\
  \bibinfo {pages} {3045} (\bibinfo {year} {2010})}\BibitemShut {NoStop}%
\bibitem [{\citenamefont {Qi}\ and\ \citenamefont {Zhang}(2011)}]{zhang-rmp}%
  \BibitemOpen
  \bibfield  {author} {\bibinfo {author} {\bibfnamefont {X.-L.}\ \bibnamefont
  {Qi}}\ and\ \bibinfo {author} {\bibfnamefont {S.-C.}\ \bibnamefont {Zhang}},\
  }\href {\doibase 10.1103/RevModPhys.83.1057} {\bibfield  {journal} {\bibinfo
  {journal} {Rev. Mod. Phys.}\ }\textbf {\bibinfo {volume} {83}},\ \bibinfo
  {pages} {1057} (\bibinfo {year} {2011})}\BibitemShut {NoStop}%
\bibitem [{\citenamefont {Armitage}\ \emph {et~al.}(2018)\citenamefont
  {Armitage}, \citenamefont {Mele},\ and\ \citenamefont
  {Vishwanath}}]{vishwanath-rmp2018}%
  \BibitemOpen
  \bibfield  {author} {\bibinfo {author} {\bibfnamefont {N.~P.}\ \bibnamefont
  {Armitage}}, \bibinfo {author} {\bibfnamefont {E.~J.}\ \bibnamefont {Mele}},
  \ and\ \bibinfo {author} {\bibfnamefont {A.}~\bibnamefont {Vishwanath}},\
  }\href {\doibase 10.1103/RevModPhys.90.015001} {\bibfield  {journal}
  {\bibinfo  {journal} {Rev. Mod. Phys.}\ }\textbf {\bibinfo {volume} {90}},\
  \bibinfo {pages} {015001} (\bibinfo {year} {2018})}\BibitemShut {NoStop}%
\bibitem [{\citenamefont {K{\"o}nig}\ \emph {et~al.}(2007)\citenamefont
  {K{\"o}nig}, \citenamefont {Wiedmann}, \citenamefont {Br{\"u}ne},
  \citenamefont {Roth}, \citenamefont {Buhmann}, \citenamefont {Molenkamp},
  \citenamefont {Qi},\ and\ \citenamefont {Zhang}}]{koenig-science2007}%
  \BibitemOpen
  \bibfield  {author} {\bibinfo {author} {\bibfnamefont {M.}~\bibnamefont
  {K{\"o}nig}}, \bibinfo {author} {\bibfnamefont {S.}~\bibnamefont {Wiedmann}},
  \bibinfo {author} {\bibfnamefont {C.}~\bibnamefont {Br{\"u}ne}}, \bibinfo
  {author} {\bibfnamefont {A.}~\bibnamefont {Roth}}, \bibinfo {author}
  {\bibfnamefont {H.}~\bibnamefont {Buhmann}}, \bibinfo {author} {\bibfnamefont
  {L.~W.}\ \bibnamefont {Molenkamp}}, \bibinfo {author} {\bibfnamefont {X.-L.}\
  \bibnamefont {Qi}}, \ and\ \bibinfo {author} {\bibfnamefont {S.-C.}\
  \bibnamefont {Zhang}},\ }\href {\doibase 10.1126/science.1148047} {\bibfield
  {journal} {\bibinfo  {journal} {Science}\ }\textbf {\bibinfo {volume}
  {318}},\ \bibinfo {pages} {766} (\bibinfo {year} {2007})}\BibitemShut
  {NoStop}%
\bibitem [{\citenamefont {Hsieh}\ \emph {et~al.}(2008)\citenamefont {Hsieh},
  \citenamefont {Qian}, \citenamefont {Wray}, \citenamefont {Xia},
  \citenamefont {Hor}, \citenamefont {Cava},\ and\ \citenamefont
  {Hasan}}]{hsieh-nature2008}%
  \BibitemOpen
  \bibfield  {author} {\bibinfo {author} {\bibfnamefont {D.}~\bibnamefont
  {Hsieh}}, \bibinfo {author} {\bibfnamefont {D.}~\bibnamefont {Qian}},
  \bibinfo {author} {\bibfnamefont {L.}~\bibnamefont {Wray}}, \bibinfo {author}
  {\bibfnamefont {Y.}~\bibnamefont {Xia}}, \bibinfo {author} {\bibfnamefont
  {Y.~S.}\ \bibnamefont {Hor}}, \bibinfo {author} {\bibfnamefont {R.~J.}\
  \bibnamefont {Cava}}, \ and\ \bibinfo {author} {\bibfnamefont {M.~Z.}\
  \bibnamefont {Hasan}},\ }\href {\doibase 10.1038/nature06843} {\bibfield
  {journal} {\bibinfo  {journal} {Nature}\ }\textbf {\bibinfo {volume} {452}},\
  \bibinfo {pages} {970} (\bibinfo {year} {2008})}\BibitemShut {NoStop}%
\bibitem [{\citenamefont {Hsieh}\ \emph {et~al.}(2009)\citenamefont {Hsieh},
  \citenamefont {Xia}, \citenamefont {Qian}, \citenamefont {Wray},
  \citenamefont {Dil}, \citenamefont {Meier}, \citenamefont {Osterwalder},
  \citenamefont {Patthey}, \citenamefont {Checkelsky}, \citenamefont {Ong}
  \emph {et~al.}}]{hsieh2009tunable}%
  \BibitemOpen
  \bibfield  {author} {\bibinfo {author} {\bibfnamefont {D.}~\bibnamefont
  {Hsieh}}, \bibinfo {author} {\bibfnamefont {Y.}~\bibnamefont {Xia}}, \bibinfo
  {author} {\bibfnamefont {D.}~\bibnamefont {Qian}}, \bibinfo {author}
  {\bibfnamefont {L.}~\bibnamefont {Wray}}, \bibinfo {author} {\bibfnamefont
  {J.}~\bibnamefont {Dil}}, \bibinfo {author} {\bibfnamefont {F.}~\bibnamefont
  {Meier}}, \bibinfo {author} {\bibfnamefont {J.}~\bibnamefont {Osterwalder}},
  \bibinfo {author} {\bibfnamefont {L.}~\bibnamefont {Patthey}}, \bibinfo
  {author} {\bibfnamefont {J.}~\bibnamefont {Checkelsky}}, \bibinfo {author}
  {\bibfnamefont {N.}~\bibnamefont {Ong}},  \emph {et~al.},\ }\href {\doibase
  10.1038/nature08234} {\bibfield  {journal} {\bibinfo  {journal} {Nature}\
  }\textbf {\bibinfo {volume} {460}},\ \bibinfo {pages} {1101} (\bibinfo {year}
  {2009})}\BibitemShut {NoStop}%
\bibitem [{\citenamefont {{Zhang}}\ \emph {et~al.}(2009)\citenamefont
  {{Zhang}}, \citenamefont {{Liu}}, \citenamefont {{Qi}}, \citenamefont
  {{Dai}}, \citenamefont {{Fang}},\ and\ \citenamefont
  {{Zhang}}}]{zhang2009TRI}%
  \BibitemOpen
  \bibfield  {author} {\bibinfo {author} {\bibfnamefont {H.}~\bibnamefont
  {{Zhang}}}, \bibinfo {author} {\bibfnamefont {C.-X.}\ \bibnamefont {{Liu}}},
  \bibinfo {author} {\bibfnamefont {X.-L.}\ \bibnamefont {{Qi}}}, \bibinfo
  {author} {\bibfnamefont {X.}~\bibnamefont {{Dai}}}, \bibinfo {author}
  {\bibfnamefont {Z.}~\bibnamefont {{Fang}}}, \ and\ \bibinfo {author}
  {\bibfnamefont {S.-C.}\ \bibnamefont {{Zhang}}},\ }\href {\doibase
  10.1038/nphys1270} {\bibfield  {journal} {\bibinfo  {journal} {Nat. Phys.}\
  }\textbf {\bibinfo {volume} {5}},\ \bibinfo {pages} {438} (\bibinfo {year}
  {2009})}\BibitemShut {NoStop}%
\bibitem [{\citenamefont {Chen}\ \emph {et~al.}(2009)\citenamefont {Chen},
  \citenamefont {Analytis}, \citenamefont {Chu}, \citenamefont {Liu},
  \citenamefont {Mo}, \citenamefont {Qi}, \citenamefont {Zhang}, \citenamefont
  {Lu}, \citenamefont {Dai}, \citenamefont {Fang}, \citenamefont {Zhang},
  \citenamefont {Fisher}, \citenamefont {Hussain},\ and\ \citenamefont
  {Shen}}]{chen-science2009}%
  \BibitemOpen
  \bibfield  {author} {\bibinfo {author} {\bibfnamefont {Y.~L.}\ \bibnamefont
  {Chen}}, \bibinfo {author} {\bibfnamefont {J.~G.}\ \bibnamefont {Analytis}},
  \bibinfo {author} {\bibfnamefont {J.-H.}\ \bibnamefont {Chu}}, \bibinfo
  {author} {\bibfnamefont {Z.~K.}\ \bibnamefont {Liu}}, \bibinfo {author}
  {\bibfnamefont {S.-K.}\ \bibnamefont {Mo}}, \bibinfo {author} {\bibfnamefont
  {X.~L.}\ \bibnamefont {Qi}}, \bibinfo {author} {\bibfnamefont {H.~J.}\
  \bibnamefont {Zhang}}, \bibinfo {author} {\bibfnamefont {D.~H.}\ \bibnamefont
  {Lu}}, \bibinfo {author} {\bibfnamefont {X.}~\bibnamefont {Dai}}, \bibinfo
  {author} {\bibfnamefont {Z.}~\bibnamefont {Fang}}, \bibinfo {author}
  {\bibfnamefont {S.~C.}\ \bibnamefont {Zhang}}, \bibinfo {author}
  {\bibfnamefont {I.~R.}\ \bibnamefont {Fisher}}, \bibinfo {author}
  {\bibfnamefont {Z.}~\bibnamefont {Hussain}}, \ and\ \bibinfo {author}
  {\bibfnamefont {Z.-X.}\ \bibnamefont {Shen}},\ }\href {\doibase
  10.1126/science.1173034} {\bibfield  {journal} {\bibinfo  {journal}
  {Science}\ }\textbf {\bibinfo {volume} {325}},\ \bibinfo {pages} {178}
  (\bibinfo {year} {2009})}\BibitemShut {NoStop}%
\bibitem [{\citenamefont {Xia}\ \emph {et~al.}(2009)\citenamefont {Xia},
  \citenamefont {Qian}, \citenamefont {Hsieh}, \citenamefont {Wray},
  \citenamefont {Pal}, \citenamefont {Lin}, \citenamefont {Bansil},
  \citenamefont {Grauer}, \citenamefont {Hor}, \citenamefont {Cava} \emph
  {et~al.}}]{xia2009observation}%
  \BibitemOpen
  \bibfield  {author} {\bibinfo {author} {\bibfnamefont {Y.}~\bibnamefont
  {Xia}}, \bibinfo {author} {\bibfnamefont {D.}~\bibnamefont {Qian}}, \bibinfo
  {author} {\bibfnamefont {D.}~\bibnamefont {Hsieh}}, \bibinfo {author}
  {\bibfnamefont {L.}~\bibnamefont {Wray}}, \bibinfo {author} {\bibfnamefont
  {A.}~\bibnamefont {Pal}}, \bibinfo {author} {\bibfnamefont {H.}~\bibnamefont
  {Lin}}, \bibinfo {author} {\bibfnamefont {A.}~\bibnamefont {Bansil}},
  \bibinfo {author} {\bibfnamefont {D.}~\bibnamefont {Grauer}}, \bibinfo
  {author} {\bibfnamefont {Y.~S.}\ \bibnamefont {Hor}}, \bibinfo {author}
  {\bibfnamefont {R.~J.}\ \bibnamefont {Cava}},  \emph {et~al.},\ }\href
  {\doibase /10.1038/nphys1274} {\bibfield  {journal} {\bibinfo  {journal}
  {Nat. Phys.}\ }\textbf {\bibinfo {volume} {5}},\ \bibinfo {pages} {398}
  (\bibinfo {year} {2009})}\BibitemShut {NoStop}%
\bibitem [{\citenamefont {Chadov}\ \emph {et~al.}(2010)\citenamefont {Chadov},
  \citenamefont {Qi}, \citenamefont {K{\"u}bler}, \citenamefont {Fecher},
  \citenamefont {Felser},\ and\ \citenamefont {Zhang}}]{felser2010natmat}%
  \BibitemOpen
  \bibfield  {author} {\bibinfo {author} {\bibfnamefont {S.}~\bibnamefont
  {Chadov}}, \bibinfo {author} {\bibfnamefont {X.}~\bibnamefont {Qi}}, \bibinfo
  {author} {\bibfnamefont {J.}~\bibnamefont {K{\"u}bler}}, \bibinfo {author}
  {\bibfnamefont {G.~H.}\ \bibnamefont {Fecher}}, \bibinfo {author}
  {\bibfnamefont {C.}~\bibnamefont {Felser}}, \ and\ \bibinfo {author}
  {\bibfnamefont {S.~C.}\ \bibnamefont {Zhang}},\ }\href {\doibase
  10.1038/nmat2770} {\bibfield  {journal} {\bibinfo  {journal} {Nat. Mater.}\
  }\textbf {\bibinfo {volume} {9}},\ \bibinfo {pages} {541} (\bibinfo {year}
  {2010})}\BibitemShut {NoStop}%
\bibitem [{\citenamefont {Lin}\ \emph {et~al.}(2010)\citenamefont {Lin},
  \citenamefont {Wray}, \citenamefont {Xia}, \citenamefont {Xu}, \citenamefont
  {Jia}, \citenamefont {Cava}, \citenamefont {Bansil},\ and\ \citenamefont
  {Hasan}}]{hasan2010natmat}%
  \BibitemOpen
  \bibfield  {author} {\bibinfo {author} {\bibfnamefont {H.}~\bibnamefont
  {Lin}}, \bibinfo {author} {\bibfnamefont {L.~A.}\ \bibnamefont {Wray}},
  \bibinfo {author} {\bibfnamefont {Y.}~\bibnamefont {Xia}}, \bibinfo {author}
  {\bibfnamefont {S.}~\bibnamefont {Xu}}, \bibinfo {author} {\bibfnamefont
  {S.}~\bibnamefont {Jia}}, \bibinfo {author} {\bibfnamefont {R.~J.}\
  \bibnamefont {Cava}}, \bibinfo {author} {\bibfnamefont {A.}~\bibnamefont
  {Bansil}}, \ and\ \bibinfo {author} {\bibfnamefont {M.~Z.}\ \bibnamefont
  {Hasan}},\ }\href {\doibase 10.1038/nmat2771} {\bibfield  {journal} {\bibinfo
   {journal} {Nat. Mater.}\ }\textbf {\bibinfo {volume} {9}},\ \bibinfo {pages}
  {546} (\bibinfo {year} {2010})}\BibitemShut {NoStop}%
\bibitem [{\citenamefont {Xiao}\ \emph {et~al.}(2010)\citenamefont {Xiao},
  \citenamefont {Yao}, \citenamefont {Feng}, \citenamefont {Wen}, \citenamefont
  {Zhu}, \citenamefont {Chen}, \citenamefont {Stocks},\ and\ \citenamefont
  {Zhang}}]{xiao2010halfheusler}%
  \BibitemOpen
  \bibfield  {author} {\bibinfo {author} {\bibfnamefont {D.}~\bibnamefont
  {Xiao}}, \bibinfo {author} {\bibfnamefont {Y.}~\bibnamefont {Yao}}, \bibinfo
  {author} {\bibfnamefont {W.}~\bibnamefont {Feng}}, \bibinfo {author}
  {\bibfnamefont {J.}~\bibnamefont {Wen}}, \bibinfo {author} {\bibfnamefont
  {W.}~\bibnamefont {Zhu}}, \bibinfo {author} {\bibfnamefont {X.-Q.}\
  \bibnamefont {Chen}}, \bibinfo {author} {\bibfnamefont {G.~M.}\ \bibnamefont
  {Stocks}}, \ and\ \bibinfo {author} {\bibfnamefont {Z.}~\bibnamefont
  {Zhang}},\ }\href {\doibase 10.1103/PhysRevLett.105.096404} {\bibfield
  {journal} {\bibinfo  {journal} {Phys. Rev. Lett.}\ }\textbf {\bibinfo
  {volume} {105}},\ \bibinfo {pages} {096404} (\bibinfo {year}
  {2010})}\BibitemShut {NoStop}%
\bibitem [{\citenamefont {Dresselhaus}(1955)}]{dresselhaus}%
  \BibitemOpen
  \bibfield  {author} {\bibinfo {author} {\bibfnamefont {G.}~\bibnamefont
  {Dresselhaus}},\ }\href {\doibase 10.1103/PhysRev.100.580} {\bibfield
  {journal} {\bibinfo  {journal} {Phys. Rev.}\ }\textbf {\bibinfo {volume}
  {100}},\ \bibinfo {pages} {580} (\bibinfo {year} {1955})}\BibitemShut
  {NoStop}%
\bibitem [{\citenamefont {Butch}\ \emph {et~al.}(2011)\citenamefont {Butch},
  \citenamefont {Syers}, \citenamefont {Kirshenbaum}, \citenamefont {Hope},\
  and\ \citenamefont {Paglione}}]{yptbi-sup-topo-prb2011}%
  \BibitemOpen
  \bibfield  {author} {\bibinfo {author} {\bibfnamefont {N.~P.}\ \bibnamefont
  {Butch}}, \bibinfo {author} {\bibfnamefont {P.}~\bibnamefont {Syers}},
  \bibinfo {author} {\bibfnamefont {K.}~\bibnamefont {Kirshenbaum}}, \bibinfo
  {author} {\bibfnamefont {A.~P.}\ \bibnamefont {Hope}}, \ and\ \bibinfo
  {author} {\bibfnamefont {J.}~\bibnamefont {Paglione}},\ }\href {\doibase
  10.1103/PhysRevB.84.220504} {\bibfield  {journal} {\bibinfo  {journal} {Phys.
  Rev. B}\ }\textbf {\bibinfo {volume} {84}},\ \bibinfo {pages} {220504}
  (\bibinfo {year} {2011})}\BibitemShut {NoStop}%
\bibitem [{\citenamefont {Tafti}\ \emph {et~al.}(2013)\citenamefont {Tafti},
  \citenamefont {Fujii}, \citenamefont {Juneau-Fecteau}, \citenamefont
  {Ren\'e~de Cotret}, \citenamefont {Doiron-Leyraud}, \citenamefont
  {Asamitsu},\ and\ \citenamefont {Taillefer}}]{topo-sup-luptbi-prb2013}%
  \BibitemOpen
  \bibfield  {author} {\bibinfo {author} {\bibfnamefont {F.~F.}\ \bibnamefont
  {Tafti}}, \bibinfo {author} {\bibfnamefont {T.}~\bibnamefont {Fujii}},
  \bibinfo {author} {\bibfnamefont {A.}~\bibnamefont {Juneau-Fecteau}},
  \bibinfo {author} {\bibfnamefont {S.}~\bibnamefont {Ren\'e~de Cotret}},
  \bibinfo {author} {\bibfnamefont {N.}~\bibnamefont {Doiron-Leyraud}},
  \bibinfo {author} {\bibfnamefont {A.}~\bibnamefont {Asamitsu}}, \ and\
  \bibinfo {author} {\bibfnamefont {L.}~\bibnamefont {Taillefer}},\ }\href
  {\doibase 10.1103/PhysRevB.87.184504} {\bibfield  {journal} {\bibinfo
  {journal} {Phys. Rev. B}\ }\textbf {\bibinfo {volume} {87}},\ \bibinfo
  {pages} {184504} (\bibinfo {year} {2013})}\BibitemShut {NoStop}%
\bibitem [{\citenamefont {Brydon}\ \emph {et~al.}(2016)\citenamefont {Brydon},
  \citenamefont {Wang}, \citenamefont {Weinert},\ and\ \citenamefont
  {Agterberg}}]{brydon-prl}%
  \BibitemOpen
  \bibfield  {author} {\bibinfo {author} {\bibfnamefont {P.~M.~R.}\
  \bibnamefont {Brydon}}, \bibinfo {author} {\bibfnamefont {L.}~\bibnamefont
  {Wang}}, \bibinfo {author} {\bibfnamefont {M.}~\bibnamefont {Weinert}}, \
  and\ \bibinfo {author} {\bibfnamefont {D.~F.}\ \bibnamefont {Agterberg}},\
  }\href {\doibase 10.1103/PhysRevLett.116.177001} {\bibfield  {journal}
  {\bibinfo  {journal} {Phys. Rev. Lett.}\ }\textbf {\bibinfo {volume} {116}},\
  \bibinfo {pages} {177001} (\bibinfo {year} {2016})}\BibitemShut {NoStop}%
\bibitem [{\citenamefont {Savary}\ \emph {et~al.}(2017)\citenamefont {Savary},
  \citenamefont {Ruhman}, \citenamefont {Venderbos}, \citenamefont {Fu},\ and\
  \citenamefont {Lee}}]{lee-prb-2017}%
  \BibitemOpen
  \bibfield  {author} {\bibinfo {author} {\bibfnamefont {L.}~\bibnamefont
  {Savary}}, \bibinfo {author} {\bibfnamefont {J.}~\bibnamefont {Ruhman}},
  \bibinfo {author} {\bibfnamefont {J.~W.~F.}\ \bibnamefont {Venderbos}},
  \bibinfo {author} {\bibfnamefont {L.}~\bibnamefont {Fu}}, \ and\ \bibinfo
  {author} {\bibfnamefont {P.~A.}\ \bibnamefont {Lee}},\ }\href {\doibase
  10.1103/PhysRevB.96.214514} {\bibfield  {journal} {\bibinfo  {journal} {Phys.
  Rev. B}\ }\textbf {\bibinfo {volume} {96}},\ \bibinfo {pages} {214514}
  (\bibinfo {year} {2017})}\BibitemShut {NoStop}%
\bibitem [{\citenamefont {Kim}\ \emph {et~al.}(2018)\citenamefont {Kim},
  \citenamefont {Wang}, \citenamefont {Nakajima}, \citenamefont {Hu},
  \citenamefont {Ziemak}, \citenamefont {Syers}, \citenamefont {Wang},
  \citenamefont {Hodovanets}, \citenamefont {Denlinger}, \citenamefont {Brydon}
  \emph {et~al.}}]{kimeaao-sci-adv}%
  \BibitemOpen
  \bibfield  {author} {\bibinfo {author} {\bibfnamefont {H.}~\bibnamefont
  {Kim}}, \bibinfo {author} {\bibfnamefont {K.}~\bibnamefont {Wang}}, \bibinfo
  {author} {\bibfnamefont {Y.}~\bibnamefont {Nakajima}}, \bibinfo {author}
  {\bibfnamefont {R.}~\bibnamefont {Hu}}, \bibinfo {author} {\bibfnamefont
  {S.}~\bibnamefont {Ziemak}}, \bibinfo {author} {\bibfnamefont
  {P.}~\bibnamefont {Syers}}, \bibinfo {author} {\bibfnamefont
  {L.}~\bibnamefont {Wang}}, \bibinfo {author} {\bibfnamefont {H.}~\bibnamefont
  {Hodovanets}}, \bibinfo {author} {\bibfnamefont {J.~D.}\ \bibnamefont
  {Denlinger}}, \bibinfo {author} {\bibfnamefont {P.~M.~R.}\ \bibnamefont
  {Brydon}},  \emph {et~al.},\ }\href {\doibase 10.1126/sciadv.aao4513}
  {\bibfield  {journal} {\bibinfo  {journal} {Sci. Adv.}\ }\textbf {\bibinfo
  {volume} {4}},\ \bibinfo {pages} {eaao4513} (\bibinfo {year}
  {2018})}\BibitemShut {NoStop}%
\bibitem [{\citenamefont {Timm}\ \emph {et~al.}(2017)\citenamefont {Timm},
  \citenamefont {Schnyder}, \citenamefont {Agterberg},\ and\ \citenamefont
  {Brydon}}]{timm-17}%
  \BibitemOpen
  \bibfield  {author} {\bibinfo {author} {\bibfnamefont {C.}~\bibnamefont
  {Timm}}, \bibinfo {author} {\bibfnamefont {A.~P.}\ \bibnamefont {Schnyder}},
  \bibinfo {author} {\bibfnamefont {D.~F.}\ \bibnamefont {Agterberg}}, \ and\
  \bibinfo {author} {\bibfnamefont {P.~M.~R.}\ \bibnamefont {Brydon}},\ }\href
  {\doibase 10.1103/PhysRevB.96.094526} {\bibfield  {journal} {\bibinfo
  {journal} {Phys. Rev. B}\ }\textbf {\bibinfo {volume} {96}},\ \bibinfo
  {pages} {094526} (\bibinfo {year} {2017})}\BibitemShut {NoStop}%
\bibitem [{\citenamefont {Venderbos}\ \emph {et~al.}(2018)\citenamefont
  {Venderbos}, \citenamefont {Savary}, \citenamefont {Ruhman}, \citenamefont
  {Lee},\ and\ \citenamefont {Fu}}]{fu-lee-prx}%
  \BibitemOpen
  \bibfield  {author} {\bibinfo {author} {\bibfnamefont {J.~W.~F.}\
  \bibnamefont {Venderbos}}, \bibinfo {author} {\bibfnamefont {L.}~\bibnamefont
  {Savary}}, \bibinfo {author} {\bibfnamefont {J.}~\bibnamefont {Ruhman}},
  \bibinfo {author} {\bibfnamefont {P.~A.}\ \bibnamefont {Lee}}, \ and\
  \bibinfo {author} {\bibfnamefont {L.}~\bibnamefont {Fu}},\ }\href {\doibase
  10.1103/PhysRevX.8.011029} {\bibfield  {journal} {\bibinfo  {journal} {Phys.
  Rev. X}\ }\textbf {\bibinfo {volume} {8}},\ \bibinfo {pages} {011029}
  (\bibinfo {year} {2018})}\BibitemShut {NoStop}%
\bibitem [{\citenamefont {Roy}\ \emph {et~al.}(2019)\citenamefont {Roy},
  \citenamefont {Ghorashi}, \citenamefont {Foster},\ and\ \citenamefont
  {Nevidomskyy}}]{ghorashi-prb19}%
  \BibitemOpen
  \bibfield  {author} {\bibinfo {author} {\bibfnamefont {B.}~\bibnamefont
  {Roy}}, \bibinfo {author} {\bibfnamefont {S.~A.~A.}\ \bibnamefont
  {Ghorashi}}, \bibinfo {author} {\bibfnamefont {M.~S.}\ \bibnamefont
  {Foster}}, \ and\ \bibinfo {author} {\bibfnamefont {A.~H.}\ \bibnamefont
  {Nevidomskyy}},\ }\href {\doibase 10.1103/PhysRevB.99.054505} {\bibfield
  {journal} {\bibinfo  {journal} {Phys. Rev. B}\ }\textbf {\bibinfo {volume}
  {99}},\ \bibinfo {pages} {054505} (\bibinfo {year} {2019})}\BibitemShut
  {NoStop}%
\bibitem [{\citenamefont {Moghaddam}\ \emph {et~al.}(2014)\citenamefont
  {Moghaddam}, \citenamefont {Kernreiter}, \citenamefont {Governale},\ and\
  \citenamefont {Z\"ulicke}}]{moghaddam14}%
  \BibitemOpen
  \bibfield  {author} {\bibinfo {author} {\bibfnamefont {A.~G.}\ \bibnamefont
  {Moghaddam}}, \bibinfo {author} {\bibfnamefont {T.}~\bibnamefont
  {Kernreiter}}, \bibinfo {author} {\bibfnamefont {M.}~\bibnamefont
  {Governale}}, \ and\ \bibinfo {author} {\bibfnamefont {U.}~\bibnamefont
  {Z\"ulicke}},\ }\href {\doibase 10.1103/PhysRevB.89.184507} {\bibfield
  {journal} {\bibinfo  {journal} {Phys. Rev. B}\ }\textbf {\bibinfo {volume}
  {89}},\ \bibinfo {pages} {184507} (\bibinfo {year} {2014})}\BibitemShut
  {NoStop}%
\bibitem [{\citenamefont {Manna}\ \emph {et~al.}(2018)\citenamefont {Manna},
  \citenamefont {Sun}, \citenamefont {Muechler}, \citenamefont {K{\"u}bler},\
  and\ \citenamefont {Felser}}]{felser2018heusler}%
  \BibitemOpen
  \bibfield  {author} {\bibinfo {author} {\bibfnamefont {K.}~\bibnamefont
  {Manna}}, \bibinfo {author} {\bibfnamefont {Y.}~\bibnamefont {Sun}}, \bibinfo
  {author} {\bibfnamefont {L.}~\bibnamefont {Muechler}}, \bibinfo {author}
  {\bibfnamefont {J.}~\bibnamefont {K{\"u}bler}}, \ and\ \bibinfo {author}
  {\bibfnamefont {C.}~\bibnamefont {Felser}},\ }\href@noop {} {\bibfield
  {journal} {\bibinfo  {journal} {Nat. Rev. Mater.}\ }\textbf {\bibinfo
  {volume} {3}},\ \bibinfo {pages} {244} (\bibinfo {year} {2018})}\BibitemShut
  {NoStop}%
\bibitem [{\citenamefont {Casper}\ \emph {et~al.}(2012)\citenamefont {Casper},
  \citenamefont {Graf}, \citenamefont {Chadov}, \citenamefont {Balke},\ and\
  \citenamefont {Felser}}]{felser-rev-energy}%
  \BibitemOpen
  \bibfield  {author} {\bibinfo {author} {\bibfnamefont {F.}~\bibnamefont
  {Casper}}, \bibinfo {author} {\bibfnamefont {T.}~\bibnamefont {Graf}},
  \bibinfo {author} {\bibfnamefont {S.}~\bibnamefont {Chadov}}, \bibinfo
  {author} {\bibfnamefont {B.}~\bibnamefont {Balke}}, \ and\ \bibinfo {author}
  {\bibfnamefont {C.}~\bibnamefont {Felser}},\ }\href {\doibase
  10.1088/0268-1242/27/6/063001} {\bibfield  {journal} {\bibinfo  {journal}
  {Semicond. Sci. Technol.}\ }\textbf {\bibinfo {volume} {27}},\ \bibinfo
  {pages} {063001} (\bibinfo {year} {2012})}\BibitemShut {NoStop}%
\bibitem [{\citenamefont {K{\"u}bler}(2017)}]{kubler2017book}%
  \BibitemOpen
  \bibfield  {author} {\bibinfo {author} {\bibfnamefont {J.}~\bibnamefont
  {K{\"u}bler}},\ }\href@noop {} {\emph {\bibinfo {title} {Theory of itinerant
  electron magnetism}}}\ (\bibinfo  {publisher} {Oxford University Press},\
  \bibinfo {year} {2017})\BibitemShut {NoStop}%
\bibitem [{\citenamefont {Ruderman}\ and\ \citenamefont
  {Kittel}(1954)}]{kittel}%
  \BibitemOpen
  \bibfield  {author} {\bibinfo {author} {\bibfnamefont {M.~A.}\ \bibnamefont
  {Ruderman}}\ and\ \bibinfo {author} {\bibfnamefont {C.}~\bibnamefont
  {Kittel}},\ }\href {\doibase 10.1103/PhysRev.96.99} {\bibfield  {journal}
  {\bibinfo  {journal} {Phys. Rev.}\ }\textbf {\bibinfo {volume} {96}},\
  \bibinfo {pages} {99} (\bibinfo {year} {1954})}\BibitemShut {NoStop}%
\bibitem [{\citenamefont {Kasuya}(1956)}]{kasuya}%
  \BibitemOpen
  \bibfield  {author} {\bibinfo {author} {\bibfnamefont {T.}~\bibnamefont
  {Kasuya}},\ }\href {\doibase 10.1143/PTP.16.45} {\bibfield  {journal}
  {\bibinfo  {journal} {Prog. Theor. Phys.}\ }\textbf {\bibinfo {volume}
  {16}},\ \bibinfo {pages} {45} (\bibinfo {year} {1956})}\BibitemShut {NoStop}%
\bibitem [{\citenamefont {Yosida}(1957)}]{yosida}%
  \BibitemOpen
  \bibfield  {author} {\bibinfo {author} {\bibfnamefont {K.}~\bibnamefont
  {Yosida}},\ }\href {\doibase 10.1103/PhysRev.106.893} {\bibfield  {journal}
  {\bibinfo  {journal} {Phys. Rev.}\ }\textbf {\bibinfo {volume} {106}},\
  \bibinfo {pages} {893} (\bibinfo {year} {1957})}\BibitemShut {NoStop}%
\bibitem [{\citenamefont {Dietl}\ and\ \citenamefont {Ohno}(2014)}]{ohno-rmp}%
  \BibitemOpen
  \bibfield  {author} {\bibinfo {author} {\bibfnamefont {T.}~\bibnamefont
  {Dietl}}\ and\ \bibinfo {author} {\bibfnamefont {H.}~\bibnamefont {Ohno}},\
  }\href {\doibase 10.1103/RevModPhys.86.187} {\bibfield  {journal} {\bibinfo
  {journal} {Rev. Mod. Phys.}\ }\textbf {\bibinfo {volume} {86}},\ \bibinfo
  {pages} {187} (\bibinfo {year} {2014})}\BibitemShut {NoStop}%
\bibitem [{\citenamefont {Ohno}(1998)}]{ohno-science}%
  \BibitemOpen
  \bibfield  {author} {\bibinfo {author} {\bibfnamefont {H.}~\bibnamefont
  {Ohno}},\ }\href {\doibase 10.1126/science.281.5379.951} {\bibfield
  {journal} {\bibinfo  {journal} {Science}\ }\textbf {\bibinfo {volume}
  {281}},\ \bibinfo {pages} {951} (\bibinfo {year} {1998})}\BibitemShut
  {NoStop}%
\bibitem [{\citenamefont {Matsukura}\ \emph {et~al.}(1998)\citenamefont
  {Matsukura}, \citenamefont {Ohno}, \citenamefont {Shen},\ and\ \citenamefont
  {Sugawara}}]{ohno-prb}%
  \BibitemOpen
  \bibfield  {author} {\bibinfo {author} {\bibfnamefont {F.}~\bibnamefont
  {Matsukura}}, \bibinfo {author} {\bibfnamefont {H.}~\bibnamefont {Ohno}},
  \bibinfo {author} {\bibfnamefont {A.}~\bibnamefont {Shen}}, \ and\ \bibinfo
  {author} {\bibfnamefont {Y.}~\bibnamefont {Sugawara}},\ }\href {\doibase
  10.1103/PhysRevB.57.R2037} {\bibfield  {journal} {\bibinfo  {journal} {Phys.
  Rev. B}\ }\textbf {\bibinfo {volume} {57}},\ \bibinfo {pages} {R2037}
  (\bibinfo {year} {1998})}\BibitemShut {NoStop}%
\bibitem [{\citenamefont {Abanin}\ and\ \citenamefont {Pesin}(2011)}]{abanin}%
  \BibitemOpen
  \bibfield  {author} {\bibinfo {author} {\bibfnamefont {D.~A.}\ \bibnamefont
  {Abanin}}\ and\ \bibinfo {author} {\bibfnamefont {D.~A.}\ \bibnamefont
  {Pesin}},\ }\href {\doibase 10.1103/PhysRevLett.106.136802} {\bibfield
  {journal} {\bibinfo  {journal} {Phys. Rev. Lett.}\ }\textbf {\bibinfo
  {volume} {106}},\ \bibinfo {pages} {136802} (\bibinfo {year}
  {2011})}\BibitemShut {NoStop}%
\bibitem [{\citenamefont {Liu}\ \emph {et~al.}(2009)\citenamefont {Liu},
  \citenamefont {Liu}, \citenamefont {Xu}, \citenamefont {Qi},\ and\
  \citenamefont {Zhang}}]{rkky-topological-zhang-2009}%
  \BibitemOpen
  \bibfield  {author} {\bibinfo {author} {\bibfnamefont {Q.}~\bibnamefont
  {Liu}}, \bibinfo {author} {\bibfnamefont {C.-X.}\ \bibnamefont {Liu}},
  \bibinfo {author} {\bibfnamefont {C.}~\bibnamefont {Xu}}, \bibinfo {author}
  {\bibfnamefont {X.-L.}\ \bibnamefont {Qi}}, \ and\ \bibinfo {author}
  {\bibfnamefont {S.-C.}\ \bibnamefont {Zhang}},\ }\href {\doibase
  10.1103/PhysRevLett.102.156603} {\bibfield  {journal} {\bibinfo  {journal}
  {Phys. Rev. Lett.}\ }\textbf {\bibinfo {volume} {102}},\ \bibinfo {pages}
  {156603} (\bibinfo {year} {2009})}\BibitemShut {NoStop}%
\bibitem [{\citenamefont {Zhu}\ \emph {et~al.}(2011)\citenamefont {Zhu},
  \citenamefont {Yao}, \citenamefont {Zhang},\ and\ \citenamefont
  {Chang}}]{chang2011}%
  \BibitemOpen
  \bibfield  {author} {\bibinfo {author} {\bibfnamefont {J.-J.}\ \bibnamefont
  {Zhu}}, \bibinfo {author} {\bibfnamefont {D.-X.}\ \bibnamefont {Yao}},
  \bibinfo {author} {\bibfnamefont {S.-C.}\ \bibnamefont {Zhang}}, \ and\
  \bibinfo {author} {\bibfnamefont {K.}~\bibnamefont {Chang}},\ }\href
  {\doibase 10.1103/PhysRevLett.106.097201} {\bibfield  {journal} {\bibinfo
  {journal} {Phys. Rev. Lett.}\ }\textbf {\bibinfo {volume} {106}},\ \bibinfo
  {pages} {097201} (\bibinfo {year} {2011})}\BibitemShut {NoStop}%
\bibitem [{\citenamefont {Hor}\ \emph {et~al.}(2010)\citenamefont {Hor},
  \citenamefont {Roushan}, \citenamefont {Beidenkopf}, \citenamefont {Seo},
  \citenamefont {Qu}, \citenamefont {Checkelsky}, \citenamefont {Wray},
  \citenamefont {Hsieh}, \citenamefont {Xia}, \citenamefont {Xu}, \citenamefont
  {Qian}, \citenamefont {Hasan}, \citenamefont {Ong}, \citenamefont {Yazdani},\
  and\ \citenamefont {Cava}}]{hor-magnetic-TI}%
  \BibitemOpen
  \bibfield  {author} {\bibinfo {author} {\bibfnamefont {Y.~S.}\ \bibnamefont
  {Hor}}, \bibinfo {author} {\bibfnamefont {P.}~\bibnamefont {Roushan}},
  \bibinfo {author} {\bibfnamefont {H.}~\bibnamefont {Beidenkopf}}, \bibinfo
  {author} {\bibfnamefont {J.}~\bibnamefont {Seo}}, \bibinfo {author}
  {\bibfnamefont {D.}~\bibnamefont {Qu}}, \bibinfo {author} {\bibfnamefont
  {J.~G.}\ \bibnamefont {Checkelsky}}, \bibinfo {author} {\bibfnamefont
  {L.~A.}\ \bibnamefont {Wray}}, \bibinfo {author} {\bibfnamefont
  {D.}~\bibnamefont {Hsieh}}, \bibinfo {author} {\bibfnamefont
  {Y.}~\bibnamefont {Xia}}, \bibinfo {author} {\bibfnamefont {S.-Y.}\
  \bibnamefont {Xu}}, \bibinfo {author} {\bibfnamefont {D.}~\bibnamefont
  {Qian}}, \bibinfo {author} {\bibfnamefont {M.~Z.}\ \bibnamefont {Hasan}},
  \bibinfo {author} {\bibfnamefont {N.~P.}\ \bibnamefont {Ong}}, \bibinfo
  {author} {\bibfnamefont {A.}~\bibnamefont {Yazdani}}, \ and\ \bibinfo
  {author} {\bibfnamefont {R.~J.}\ \bibnamefont {Cava}},\ }\href {\doibase
  10.1103/PhysRevB.81.195203} {\bibfield  {journal} {\bibinfo  {journal} {Phys.
  Rev. B}\ }\textbf {\bibinfo {volume} {81}},\ \bibinfo {pages} {195203}
  (\bibinfo {year} {2010})}\BibitemShut {NoStop}%
\bibitem [{\citenamefont {Checkelsky}\ \emph {et~al.}(2012)\citenamefont
  {Checkelsky}, \citenamefont {Ye}, \citenamefont {Onose}, \citenamefont
  {Iwasa},\ and\ \citenamefont {Tokura}}]{checkelsky2012dirac}%
  \BibitemOpen
  \bibfield  {author} {\bibinfo {author} {\bibfnamefont {J.~G.}\ \bibnamefont
  {Checkelsky}}, \bibinfo {author} {\bibfnamefont {J.}~\bibnamefont {Ye}},
  \bibinfo {author} {\bibfnamefont {Y.}~\bibnamefont {Onose}}, \bibinfo
  {author} {\bibfnamefont {Y.}~\bibnamefont {Iwasa}}, \ and\ \bibinfo {author}
  {\bibfnamefont {Y.}~\bibnamefont {Tokura}},\ }\href {\doibase
  10.1038/nphys2388} {\bibfield  {journal} {\bibinfo  {journal} {Nat. Phys.}\
  }\textbf {\bibinfo {volume} {8}},\ \bibinfo {pages} {729} (\bibinfo {year}
  {2012})}\BibitemShut {NoStop}%
\bibitem [{\citenamefont {Yafet}(1987)}]{rkky-1d-yafet}%
  \BibitemOpen
  \bibfield  {author} {\bibinfo {author} {\bibfnamefont {Y.}~\bibnamefont
  {Yafet}},\ }\href {\doibase 10.1103/PhysRevB.36.3948} {\bibfield  {journal}
  {\bibinfo  {journal} {Phys. Rev. B}\ }\textbf {\bibinfo {volume} {36}},\
  \bibinfo {pages} {3948} (\bibinfo {year} {1987})}\BibitemShut {NoStop}%
\bibitem [{\citenamefont {B\'eal-Monod}(1987)}]{rkky-2d-1987}%
  \BibitemOpen
  \bibfield  {author} {\bibinfo {author} {\bibfnamefont {M.~T.}\ \bibnamefont
  {B\'eal-Monod}},\ }\href {\doibase 10.1103/PhysRevB.36.8835} {\bibfield
  {journal} {\bibinfo  {journal} {Phys. Rev. B}\ }\textbf {\bibinfo {volume}
  {36}},\ \bibinfo {pages} {8835} (\bibinfo {year} {1987})}\BibitemShut
  {NoStop}%
\bibitem [{\citenamefont {Litvinov}\ and\ \citenamefont
  {Dugaev}(1998)}]{rkky-1d-2d-dugaev}%
  \BibitemOpen
  \bibfield  {author} {\bibinfo {author} {\bibfnamefont {V.~I.}\ \bibnamefont
  {Litvinov}}\ and\ \bibinfo {author} {\bibfnamefont {V.~K.}\ \bibnamefont
  {Dugaev}},\ }\href {\doibase 10.1103/PhysRevB.58.3584} {\bibfield  {journal}
  {\bibinfo  {journal} {Phys. Rev. B}\ }\textbf {\bibinfo {volume} {58}},\
  \bibinfo {pages} {3584} (\bibinfo {year} {1998})}\BibitemShut {NoStop}%
\bibitem [{\citenamefont {Aristov}(1997)}]{rkky-prb-any-d}%
  \BibitemOpen
  \bibfield  {author} {\bibinfo {author} {\bibfnamefont {D.~N.}\ \bibnamefont
  {Aristov}},\ }\href {\doibase 10.1103/PhysRevB.55.8064} {\bibfield  {journal}
  {\bibinfo  {journal} {Phys. Rev. B}\ }\textbf {\bibinfo {volume} {55}},\
  \bibinfo {pages} {8064} (\bibinfo {year} {1997})}\BibitemShut {NoStop}%
\bibitem [{\citenamefont {Black-Schaffer}(2010)}]{rkky-graphene-2010}%
  \BibitemOpen
  \bibfield  {author} {\bibinfo {author} {\bibfnamefont {A.~M.}\ \bibnamefont
  {Black-Schaffer}},\ }\href {\doibase 10.1103/PhysRevB.81.205416} {\bibfield
  {journal} {\bibinfo  {journal} {Phys. Rev. B}\ }\textbf {\bibinfo {volume}
  {81}},\ \bibinfo {pages} {205416} (\bibinfo {year} {2010})}\BibitemShut
  {NoStop}%
\bibitem [{\citenamefont {Sherafati}\ and\ \citenamefont
  {Satpathy}(2011)}]{rkky-graphene-2011}%
  \BibitemOpen
  \bibfield  {author} {\bibinfo {author} {\bibfnamefont {M.}~\bibnamefont
  {Sherafati}}\ and\ \bibinfo {author} {\bibfnamefont {S.}~\bibnamefont
  {Satpathy}},\ }\href {\doibase 10.1103/PhysRevB.83.165425} {\bibfield
  {journal} {\bibinfo  {journal} {Phys. Rev. B}\ }\textbf {\bibinfo {volume}
  {83}},\ \bibinfo {pages} {165425} (\bibinfo {year} {2011})}\BibitemShut
  {NoStop}%
\bibitem [{\citenamefont {Kogan}(2011)}]{rkky-kogan}%
  \BibitemOpen
  \bibfield  {author} {\bibinfo {author} {\bibfnamefont {E.}~\bibnamefont
  {Kogan}},\ }\href {\doibase 10.1103/PhysRevB.84.115119} {\bibfield  {journal}
  {\bibinfo  {journal} {Phys. Rev. B}\ }\textbf {\bibinfo {volume} {84}},\
  \bibinfo {pages} {115119} (\bibinfo {year} {2011})}\BibitemShut {NoStop}%
\bibitem [{\citenamefont {Chang}\ \emph {et~al.}(2015)\citenamefont {Chang},
  \citenamefont {Zhou}, \citenamefont {Wang}, \citenamefont {Shan},\ and\
  \citenamefont {Xiao}}]{rkky-weyl}%
  \BibitemOpen
  \bibfield  {author} {\bibinfo {author} {\bibfnamefont {H.-R.}\ \bibnamefont
  {Chang}}, \bibinfo {author} {\bibfnamefont {J.}~\bibnamefont {Zhou}},
  \bibinfo {author} {\bibfnamefont {S.-X.}\ \bibnamefont {Wang}}, \bibinfo
  {author} {\bibfnamefont {W.-Y.}\ \bibnamefont {Shan}}, \ and\ \bibinfo
  {author} {\bibfnamefont {D.}~\bibnamefont {Xiao}},\ }\href {\doibase
  10.1103/PhysRevB.92.241103} {\bibfield  {journal} {\bibinfo  {journal} {Phys.
  Rev. B}\ }\textbf {\bibinfo {volume} {92}},\ \bibinfo {pages} {241103}
  (\bibinfo {year} {2015})}\BibitemShut {NoStop}%
\bibitem [{\citenamefont {Hosseini}\ and\ \citenamefont
  {Askari}(2015)}]{hosseini}%
  \BibitemOpen
  \bibfield  {author} {\bibinfo {author} {\bibfnamefont {M.~V.}\ \bibnamefont
  {Hosseini}}\ and\ \bibinfo {author} {\bibfnamefont {M.}~\bibnamefont
  {Askari}},\ }\href {\doibase 10.1103/PhysRevB.92.224435} {\bibfield
  {journal} {\bibinfo  {journal} {Phys. Rev. B}\ }\textbf {\bibinfo {volume}
  {92}},\ \bibinfo {pages} {224435} (\bibinfo {year} {2015})}\BibitemShut
  {NoStop}%
\bibitem [{\citenamefont {Akbari}\ \emph {et~al.}(2011)\citenamefont {Akbari},
  \citenamefont {Eremin},\ and\ \citenamefont {Thalmeier}}]{rkky-akbari}%
  \BibitemOpen
  \bibfield  {author} {\bibinfo {author} {\bibfnamefont {A.}~\bibnamefont
  {Akbari}}, \bibinfo {author} {\bibfnamefont {I.}~\bibnamefont {Eremin}}, \
  and\ \bibinfo {author} {\bibfnamefont {P.}~\bibnamefont {Thalmeier}},\ }\href
  {\doibase 10.1103/PhysRevB.84.134513} {\bibfield  {journal} {\bibinfo
  {journal} {Phys. Rev. B}\ }\textbf {\bibinfo {volume} {84}},\ \bibinfo
  {pages} {134513} (\bibinfo {year} {2011})}\BibitemShut {NoStop}%
\bibitem [{\citenamefont {Imamura}\ \emph {et~al.}(2004)\citenamefont
  {Imamura}, \citenamefont {Bruno},\ and\ \citenamefont
  {Utsumi}}]{bruno-prb-2004}%
  \BibitemOpen
  \bibfield  {author} {\bibinfo {author} {\bibfnamefont {H.}~\bibnamefont
  {Imamura}}, \bibinfo {author} {\bibfnamefont {P.}~\bibnamefont {Bruno}}, \
  and\ \bibinfo {author} {\bibfnamefont {Y.}~\bibnamefont {Utsumi}},\ }\href
  {\doibase 10.1103/PhysRevB.69.121303} {\bibfield  {journal} {\bibinfo
  {journal} {Phys. Rev. B}\ }\textbf {\bibinfo {volume} {69}},\ \bibinfo
  {pages} {121303} (\bibinfo {year} {2004})}\BibitemShut {NoStop}%
\bibitem [{\citenamefont {Lyu}\ \emph {et~al.}(2007)\citenamefont {Lyu},
  \citenamefont {Liu},\ and\ \citenamefont {Zhang}}]{chao2007}%
  \BibitemOpen
  \bibfield  {author} {\bibinfo {author} {\bibfnamefont {P.}~\bibnamefont
  {Lyu}}, \bibinfo {author} {\bibfnamefont {N.-N.}\ \bibnamefont {Liu}}, \ and\
  \bibinfo {author} {\bibfnamefont {C.}~\bibnamefont {Zhang}},\ }\href
  {\doibase 10.1063/1.2817405} {\bibfield  {journal} {\bibinfo  {journal} {J.
  App. Phys.}\ }\textbf {\bibinfo {volume} {102}},\ \bibinfo {pages} {103910}
  (\bibinfo {year} {2007})}\BibitemShut {NoStop}%
\bibitem [{\citenamefont {Simon}\ \emph {et~al.}(2008)\citenamefont {Simon},
  \citenamefont {Braunecker},\ and\ \citenamefont {Loss}}]{loss2008}%
  \BibitemOpen
  \bibfield  {author} {\bibinfo {author} {\bibfnamefont {P.}~\bibnamefont
  {Simon}}, \bibinfo {author} {\bibfnamefont {B.}~\bibnamefont {Braunecker}}, \
  and\ \bibinfo {author} {\bibfnamefont {D.}~\bibnamefont {Loss}},\ }\href
  {\doibase 10.1103/PhysRevB.77.045108} {\bibfield  {journal} {\bibinfo
  {journal} {Phys. Rev. B}\ }\textbf {\bibinfo {volume} {77}},\ \bibinfo
  {pages} {045108} (\bibinfo {year} {2008})}\BibitemShut {NoStop}%
\bibitem [{\citenamefont {Schulz}\ \emph {et~al.}(2009)\citenamefont {Schulz},
  \citenamefont {De~Martino}, \citenamefont {Ingenhoven},\ and\ \citenamefont
  {Egger}}]{egger2009}%
  \BibitemOpen
  \bibfield  {author} {\bibinfo {author} {\bibfnamefont {A.}~\bibnamefont
  {Schulz}}, \bibinfo {author} {\bibfnamefont {A.}~\bibnamefont {De~Martino}},
  \bibinfo {author} {\bibfnamefont {P.}~\bibnamefont {Ingenhoven}}, \ and\
  \bibinfo {author} {\bibfnamefont {R.}~\bibnamefont {Egger}},\ }\href
  {\doibase 10.1103/PhysRevB.79.205432} {\bibfield  {journal} {\bibinfo
  {journal} {Phys. Rev. B}\ }\textbf {\bibinfo {volume} {79}},\ \bibinfo
  {pages} {205432} (\bibinfo {year} {2009})}\BibitemShut {NoStop}%
\bibitem [{\citenamefont {Zhu}\ \emph {et~al.}(2010)\citenamefont {Zhu},
  \citenamefont {Chang}, \citenamefont {Liu},\ and\ \citenamefont
  {Lin}}]{jiaji2010}%
  \BibitemOpen
  \bibfield  {author} {\bibinfo {author} {\bibfnamefont {J.-J.}\ \bibnamefont
  {Zhu}}, \bibinfo {author} {\bibfnamefont {K.}~\bibnamefont {Chang}}, \bibinfo
  {author} {\bibfnamefont {R.-B.}\ \bibnamefont {Liu}}, \ and\ \bibinfo
  {author} {\bibfnamefont {H.-Q.}\ \bibnamefont {Lin}},\ }\href {\doibase
  10.1103/PhysRevB.81.113302} {\bibfield  {journal} {\bibinfo  {journal} {Phys.
  Rev. B}\ }\textbf {\bibinfo {volume} {81}},\ \bibinfo {pages} {113302}
  (\bibinfo {year} {2010})}\BibitemShut {NoStop}%
\bibitem [{\citenamefont {Klinovaja}\ and\ \citenamefont
  {Loss}(2013)}]{loss2013}%
  \BibitemOpen
  \bibfield  {author} {\bibinfo {author} {\bibfnamefont {J.}~\bibnamefont
  {Klinovaja}}\ and\ \bibinfo {author} {\bibfnamefont {D.}~\bibnamefont
  {Loss}},\ }\href {\doibase 10.1103/PhysRevB.87.045422} {\bibfield  {journal}
  {\bibinfo  {journal} {Phys. Rev. B}\ }\textbf {\bibinfo {volume} {87}},\
  \bibinfo {pages} {045422} (\bibinfo {year} {2013})}\BibitemShut {NoStop}%
\bibitem [{\citenamefont {Wang}\ \emph {et~al.}(2017)\citenamefont {Wang},
  \citenamefont {Chang},\ and\ \citenamefont {Zhou}}]{rkky-shiba-3d}%
  \BibitemOpen
  \bibfield  {author} {\bibinfo {author} {\bibfnamefont {S.-X.}\ \bibnamefont
  {Wang}}, \bibinfo {author} {\bibfnamefont {H.-R.}\ \bibnamefont {Chang}}, \
  and\ \bibinfo {author} {\bibfnamefont {J.}~\bibnamefont {Zhou}},\ }\href
  {\doibase 10.1103/PhysRevB.96.115204} {\bibfield  {journal} {\bibinfo
  {journal} {Phys. Rev. B}\ }\textbf {\bibinfo {volume} {96}},\ \bibinfo
  {pages} {115204} (\bibinfo {year} {2017})}\BibitemShut {NoStop}%
\bibitem [{\citenamefont {Asmar}\ and\ \citenamefont
  {Tse}(2019)}]{tse-bulk-rashba}%
  \BibitemOpen
  \bibfield  {author} {\bibinfo {author} {\bibfnamefont {M.~M.}\ \bibnamefont
  {Asmar}}\ and\ \bibinfo {author} {\bibfnamefont {W.-K.}\ \bibnamefont
  {Tse}},\ }\href {\doibase 10.1103/PhysRevB.100.014410} {\bibfield  {journal}
  {\bibinfo  {journal} {Phys. Rev. B}\ }\textbf {\bibinfo {volume} {100}},\
  \bibinfo {pages} {014410} (\bibinfo {year} {2019})}\BibitemShut {NoStop}%
\bibitem [{\citenamefont {Paul}\ \emph {et~al.}(2019)\citenamefont {Paul},
  \citenamefont {Islam},\ and\ \citenamefont {Saha}}]{saha2019}%
  \BibitemOpen
  \bibfield  {author} {\bibinfo {author} {\bibfnamefont {G.~C.}\ \bibnamefont
  {Paul}}, \bibinfo {author} {\bibfnamefont {S.~F.}\ \bibnamefont {Islam}}, \
  and\ \bibinfo {author} {\bibfnamefont {A.}~\bibnamefont {Saha}},\ }\href
  {\doibase 10.1103/PhysRevB.99.155418} {\bibfield  {journal} {\bibinfo
  {journal} {Phys. Rev. B}\ }\textbf {\bibinfo {volume} {99}},\ \bibinfo
  {pages} {155418} (\bibinfo {year} {2019})}\BibitemShut {NoStop}%
\bibitem [{\citenamefont {Klinovaja}\ \emph {et~al.}(2013)\citenamefont
  {Klinovaja}, \citenamefont {Stano}, \citenamefont {Yazdani},\ and\
  \citenamefont {Loss}}]{loss-mzm}%
  \BibitemOpen
  \bibfield  {author} {\bibinfo {author} {\bibfnamefont {J.}~\bibnamefont
  {Klinovaja}}, \bibinfo {author} {\bibfnamefont {P.}~\bibnamefont {Stano}},
  \bibinfo {author} {\bibfnamefont {A.}~\bibnamefont {Yazdani}}, \ and\
  \bibinfo {author} {\bibfnamefont {D.}~\bibnamefont {Loss}},\ }\href {\doibase
  10.1103/PhysRevLett.111.186805} {\bibfield  {journal} {\bibinfo  {journal}
  {Phys. Rev. Lett.}\ }\textbf {\bibinfo {volume} {111}},\ \bibinfo {pages}
  {186805} (\bibinfo {year} {2013})}\BibitemShut {NoStop}%
\bibitem [{\citenamefont {Braunecker}\ and\ \citenamefont
  {Simon}(2013)}]{simon-2013}%
  \BibitemOpen
  \bibfield  {author} {\bibinfo {author} {\bibfnamefont {B.}~\bibnamefont
  {Braunecker}}\ and\ \bibinfo {author} {\bibfnamefont {P.}~\bibnamefont
  {Simon}},\ }\href {\doibase 10.1103/PhysRevLett.111.147202} {\bibfield
  {journal} {\bibinfo  {journal} {Phys. Rev. Lett.}\ }\textbf {\bibinfo
  {volume} {111}},\ \bibinfo {pages} {147202} (\bibinfo {year}
  {2013})}\BibitemShut {NoStop}%
\bibitem [{\citenamefont {Pientka}\ \emph {et~al.}(2013)\citenamefont
  {Pientka}, \citenamefont {Glazman},\ and\ \citenamefont {von
  Oppen}}]{glazman-zmz}%
  \BibitemOpen
  \bibfield  {author} {\bibinfo {author} {\bibfnamefont {F.}~\bibnamefont
  {Pientka}}, \bibinfo {author} {\bibfnamefont {L.~I.}\ \bibnamefont
  {Glazman}}, \ and\ \bibinfo {author} {\bibfnamefont {F.}~\bibnamefont {von
  Oppen}},\ }\href {\doibase 10.1103/PhysRevB.88.155420} {\bibfield  {journal}
  {\bibinfo  {journal} {Phys. Rev. B}\ }\textbf {\bibinfo {volume} {88}},\
  \bibinfo {pages} {155420} (\bibinfo {year} {2013})}\BibitemShut {NoStop}%
\bibitem [{\citenamefont {Kim}\ \emph {et~al.}(2014)\citenamefont {Kim},
  \citenamefont {Cheng}, \citenamefont {Bauer}, \citenamefont {Lutchyn},\ and\
  \citenamefont {Das~Sarma}}]{sarma-mzm}%
  \BibitemOpen
  \bibfield  {author} {\bibinfo {author} {\bibfnamefont {Y.}~\bibnamefont
  {Kim}}, \bibinfo {author} {\bibfnamefont {M.}~\bibnamefont {Cheng}}, \bibinfo
  {author} {\bibfnamefont {B.}~\bibnamefont {Bauer}}, \bibinfo {author}
  {\bibfnamefont {R.~M.}\ \bibnamefont {Lutchyn}}, \ and\ \bibinfo {author}
  {\bibfnamefont {S.}~\bibnamefont {Das~Sarma}},\ }\href {\doibase
  10.1103/PhysRevB.90.060401} {\bibfield  {journal} {\bibinfo  {journal} {Phys.
  Rev. B}\ }\textbf {\bibinfo {volume} {90}},\ \bibinfo {pages} {060401}
  (\bibinfo {year} {2014})}\BibitemShut {NoStop}%
\bibitem [{\citenamefont {P\"oyh\"onen}\ \emph {et~al.}(2014)\citenamefont
  {P\"oyh\"onen}, \citenamefont {Weststr\"om}, \citenamefont {R\"ontynen},\
  and\ \citenamefont {Ojanen}}]{ojanen}%
  \BibitemOpen
  \bibfield  {author} {\bibinfo {author} {\bibfnamefont {K.}~\bibnamefont
  {P\"oyh\"onen}}, \bibinfo {author} {\bibfnamefont {A.}~\bibnamefont
  {Weststr\"om}}, \bibinfo {author} {\bibfnamefont {J.}~\bibnamefont
  {R\"ontynen}}, \ and\ \bibinfo {author} {\bibfnamefont {T.}~\bibnamefont
  {Ojanen}},\ }\href {\doibase 10.1103/PhysRevB.89.115109} {\bibfield
  {journal} {\bibinfo  {journal} {Phys. Rev. B}\ }\textbf {\bibinfo {volume}
  {89}},\ \bibinfo {pages} {115109} (\bibinfo {year} {2014})}\BibitemShut
  {NoStop}%
\bibitem [{\citenamefont {Coqblin}\ and\ \citenamefont
  {Schrieffer}(1969)}]{schrieffer69}%
  \BibitemOpen
  \bibfield  {author} {\bibinfo {author} {\bibfnamefont {B.}~\bibnamefont
  {Coqblin}}\ and\ \bibinfo {author} {\bibfnamefont {J.~R.}\ \bibnamefont
  {Schrieffer}},\ }\href {\doibase 10.1103/PhysRev.185.847} {\bibfield
  {journal} {\bibinfo  {journal} {Phys. Rev.}\ }\textbf {\bibinfo {volume}
  {185}},\ \bibinfo {pages} {847} (\bibinfo {year} {1969})}\BibitemShut
  {NoStop}%
\bibitem [{\citenamefont {Dzyaloshinsky}(1958)}]{dzyaloshinsky1958}%
  \BibitemOpen
  \bibfield  {author} {\bibinfo {author} {\bibfnamefont {I.}~\bibnamefont
  {Dzyaloshinsky}},\ }\href {\doibase
  https://doi.org/10.1016/0022-3697(58)90076-3} {\bibfield  {journal} {\bibinfo
   {journal} {J. Phys. Chem. Solids}\ }\textbf {\bibinfo {volume} {4}},\
  \bibinfo {pages} {241} (\bibinfo {year} {1958})}\BibitemShut {NoStop}%
\bibitem [{\citenamefont {Moriya}(1960{\natexlab{a}})}]{moriya-prl-1960}%
  \BibitemOpen
  \bibfield  {author} {\bibinfo {author} {\bibfnamefont {T.}~\bibnamefont
  {Moriya}},\ }\href {\doibase 10.1103/PhysRevLett.4.228} {\bibfield  {journal}
  {\bibinfo  {journal} {Phys. Rev. Lett.}\ }\textbf {\bibinfo {volume} {4}},\
  \bibinfo {pages} {228} (\bibinfo {year} {1960}{\natexlab{a}})}\BibitemShut
  {NoStop}%
\bibitem [{\citenamefont {Moriya}(1960{\natexlab{b}})}]{moriya-pr-1960}%
  \BibitemOpen
  \bibfield  {author} {\bibinfo {author} {\bibfnamefont {T.}~\bibnamefont
  {Moriya}},\ }\href {\doibase 10.1103/PhysRev.120.91} {\bibfield  {journal}
  {\bibinfo  {journal} {Phys. Rev.}\ }\textbf {\bibinfo {volume} {120}},\
  \bibinfo {pages} {91} (\bibinfo {year} {1960}{\natexlab{b}})}\BibitemShut
  {NoStop}%
\bibitem [{\citenamefont {S{\'o}lyom}(2008)}]{solyom2008fundamentals}%
  \BibitemOpen
  \bibfield  {author} {\bibinfo {author} {\bibfnamefont {J.}~\bibnamefont
  {S{\'o}lyom}},\ }\href@noop {} {\emph {\bibinfo {title} {Fundamentals of the
  Physics of Solids: Volume II: Electronic Properties}}},\ Vol.~\bibinfo
  {volume} {2}\ (\bibinfo  {publisher} {Springer Science \& Business Media},\
  \bibinfo {year} {2008})\BibitemShut {NoStop}%
\bibitem [{\citenamefont {Kernreiter}(2013)}]{kernreiter13}%
  \BibitemOpen
  \bibfield  {author} {\bibinfo {author} {\bibfnamefont {T.}~\bibnamefont
  {Kernreiter}},\ }\href {\doibase 10.1103/PhysRevB.88.085417} {\bibfield
  {journal} {\bibinfo  {journal} {Phys. Rev. B}\ }\textbf {\bibinfo {volume}
  {88}},\ \bibinfo {pages} {085417} (\bibinfo {year} {2013})}\BibitemShut
  {NoStop}%
\bibitem [{\citenamefont {Verma}\ \emph {et~al.}(2019)\citenamefont {Verma},
  \citenamefont {Kundu},\ and\ \citenamefont {Ghosh}}]{verma19}%
  \BibitemOpen
  \bibfield  {author} {\bibinfo {author} {\bibfnamefont {S.}~\bibnamefont
  {Verma}}, \bibinfo {author} {\bibfnamefont {A.}~\bibnamefont {Kundu}}, \ and\
  \bibinfo {author} {\bibfnamefont {T.~K.}\ \bibnamefont {Ghosh}},\ }\href
  {\doibase 10.1063/1.5097673} {\bibfield  {journal} {\bibinfo  {journal} {J.
  Appl. Phys.}\ }\textbf {\bibinfo {volume} {125}},\ \bibinfo {pages} {233903}
  (\bibinfo {year} {2019})}\BibitemShut {NoStop}%
\bibitem [{\citenamefont {Akbari}\ \emph {et~al.}(2013)\citenamefont {Akbari},
  \citenamefont {Thalmeier},\ and\ \citenamefont {Eremin}}]{Akbari-2013}%
  \BibitemOpen
  \bibfield  {author} {\bibinfo {author} {\bibfnamefont {A.}~\bibnamefont
  {Akbari}}, \bibinfo {author} {\bibfnamefont {P.}~\bibnamefont {Thalmeier}}, \
  and\ \bibinfo {author} {\bibfnamefont {I.}~\bibnamefont {Eremin}},\ }\href
  {\doibase 10.1088/1367-2630/15/3/033034} {\bibfield  {journal} {\bibinfo
  {journal} {New J. Phys.}\ }\textbf {\bibinfo {volume} {15}},\ \bibinfo
  {pages} {033034} (\bibinfo {year} {2013})}\BibitemShut {NoStop}%
\bibitem [{\citenamefont {Asmar}\ \emph {et~al.}(2017)\citenamefont {Asmar},
  \citenamefont {Sheehy},\ and\ \citenamefont {Vekhter}}]{asmar-2017}%
  \BibitemOpen
  \bibfield  {author} {\bibinfo {author} {\bibfnamefont {M.~M.}\ \bibnamefont
  {Asmar}}, \bibinfo {author} {\bibfnamefont {D.~E.}\ \bibnamefont {Sheehy}}, \
  and\ \bibinfo {author} {\bibfnamefont {I.}~\bibnamefont {Vekhter}},\ }\href
  {\doibase 10.1103/PhysRevB.95.241115} {\bibfield  {journal} {\bibinfo
  {journal} {Phys. Rev. B}\ }\textbf {\bibinfo {volume} {95}},\ \bibinfo
  {pages} {241115} (\bibinfo {year} {2017})}\BibitemShut {NoStop}%
\end{thebibliography}%
\end{document}